# Upcycling Human Excrement: The Gut Microbiome to Soil Microbiome Axis

Jeff Meilander[1,2], Chloe Herman[1,3], Andrew Manley[1,2], Georgia Augustine[2], Dawn Birdsell[6], Evan Bolyen[1], Kimberly R. Celona[6], Hayden Coffey[2], Jill Cocking[4], Teddy Donoghue[2], Alexis Draves[2], Daryn Erickson[2,5,6], Marissa Foley[2], Liz Gehret[1], Johannah Hagen[1], Crystal Hepp[3,5], Parker Ingram[2], David John[2], Katarina Kadar[2], Paul Keim[1], Victoria Lloyd[2], Christina Osterink[2], Victoria Queeney[2], Diego Ramirez[2], Antonio Romero[2], Megan C. Ruby[1], Jason W. Sahl[2,5], Sydni Soloway[1], Nathan E. Stone[6], Shannon Trottier[2], Kaleb Van Orden[2], Alexis Painter[2], Sam Wallace[2], Larissa Wilcox[2], Colin V. Wood[1], Jaiden Yancey[2], and J. Gregory Caporaso[1,2,3,5]

1. Center for Applied Microbiome Science, Pathogen and Microbiome Institute, Northern Arizona University, Flagstaff, AZ, USA.
2. Department of Biological Sciences, Northern Arizona University, Flagstaff, AZ, USA.
3. School of Informatics, Computing, and Cyber Systems, Northern Arizona University, Flagstaff, AZ, USA
4. Genetics Core Facility, Office of the Vice President for Research, Northern Arizona University, Flagstaff, AZ, USA
5. Pathogen and Microbiome Division, Translational Genomics Research Institute, Flagstaff, AZ, USA
6. Pathogen and Microbiome Institute, Northern Arizona University, Flagstaff, AZ USA

**Human excrement composting (HEC) is a sustainable strategy for human excrement (HE) management that recycles nutrients and mitigates health risks while reducing reliance on freshwater, fossil fuels, and fertilizers. We present a comprehensive microbial time series analysis of HEC and show that the initial gut-like microbiome of HEC systems transitions to a microbiome similar to soil and traditional compost in fifteen biological replicates tracked weekly for one year.**

Traditional wastewater treatment infrastructure depletes freshwater, diverts nutrients to landfills, requires significant amounts of energy, and fails to meet the needs of over 2 billion people[1] resulting in public and environmental health concerns. Human excrement (HE) management represents a critical global challenge, particularly in the context of exponential population growth and rapidly changing climates.

Human excrement composting (HEC) utilizing composting toilets (CTs) presents a potential solution through low-cost facilities for defecation, eliminating flushing, supporting independent HE and industrial waste streams, and reducing the burden on wastewater treatment plants (WWTPs). Additionally, installing CTs creates economic opportunities and recycles nutrients[2].

Extensive research has been conducted on microbial dynamics of thermophilic composting systems[3], however, microbial succession during mesophilic composting of HE remains understudied and is essential for contextualizing thermophilic composting systems. By sequencing the V4 region of the 16S rRNA gene coupled with qPCR and culturing experiments, this study investigated the microbial succession during 1 year of mesophilic HEC across 15 biological replicates revealing the gut microbiome to soil microbiome (gut-to-soil) axis in the highest resolution to date (Extended Data Fig. 1 and Methods). We hypothesized that (1) the microbiome of 1 year old mesophilic compost would resemble that of a soil and/or food and landscape waste compost (FLWC) microbiome more closely than the original HE (Fig. 1); and (2) human fecal indicators, *E. coli* and *C. perfringens*, would be undetectable after 52 weeks (Fig. 2). Each replicate was maintained within a modified 19-liter bucket for the experiment (Methods), therefore we reference the replicates as "buckets" throughout this text.

Microbial communities driving efficient composting operate within ecological niches defined by optimal ranges of variables including temperature (>55°C - thermophilic range), carbon to nitrogen ratio (C:N 30:1), and moisture content (45-65%)[4]. Optimizing these allows microbiota to accelerate the biodegradation of recalcitrant compounds, enhance humification and reduce risks associated with pathogens, "forever chemicals" (PFAS), pharmaceuticals, and antimicrobial resistance genes (ARG)[5].

"Thermophilic temperatures" are preferred to effectively sanitize compost and accelerate decomposition[4]. HEC temperatures were tracked with seasonal and ambient temperatures (Extended Data Fig. 2a; Supplementary Text 2). A PERMANOVA analysis demonstrated HEC microbiome variation with season ($R^2 = 0.12$, $p < 0.001$) and bucket ($R^2 = 0.05$, $p < 0.001$). The interaction between season and bucket was also significant ($R^2 = 0.072$, $p < 0.001$), suggesting that the influence of season is different in each bucket. Residual variability was high ($R^2 = 0.762$), suggesting that considerable variability in microbial composition is unexplained by these variables.

The preferred composting pH occurs between 6.5 and 8.0. Composting time point 1 (TP-1) exhibited pH values ranging from 5.5-10 across buckets decreasing to a range of 4.5-6.5 by TP-52 (Extended Data Fig. 2b,c and Supplementary Text 2) likely attributed to the accumulation of organic acids and $CO_2$ dissolution[6].

Ordination of Unweighted UniFrac distances indicated microbiome compositions transitioning from HE microbiomes towards reference sample microbiomes along Axis 1. Spearman's correlation analysis revealed a statistically significant, positive correlation between Axis 1 and time ($\rho = 0.431$; $p = 0.001$) and pH ($\rho = 0.336$; $p = 0.001$) for HEC samples.

Buckets 1, 4, 5, and 7 (B1, B4, B5, B7) demonstrate clear differences in the extent, rate, and trajectory of microbiome transitions from HE-like towards soil and FLWC reference samples (Fig. 1 and Extended Data Fig. 3).

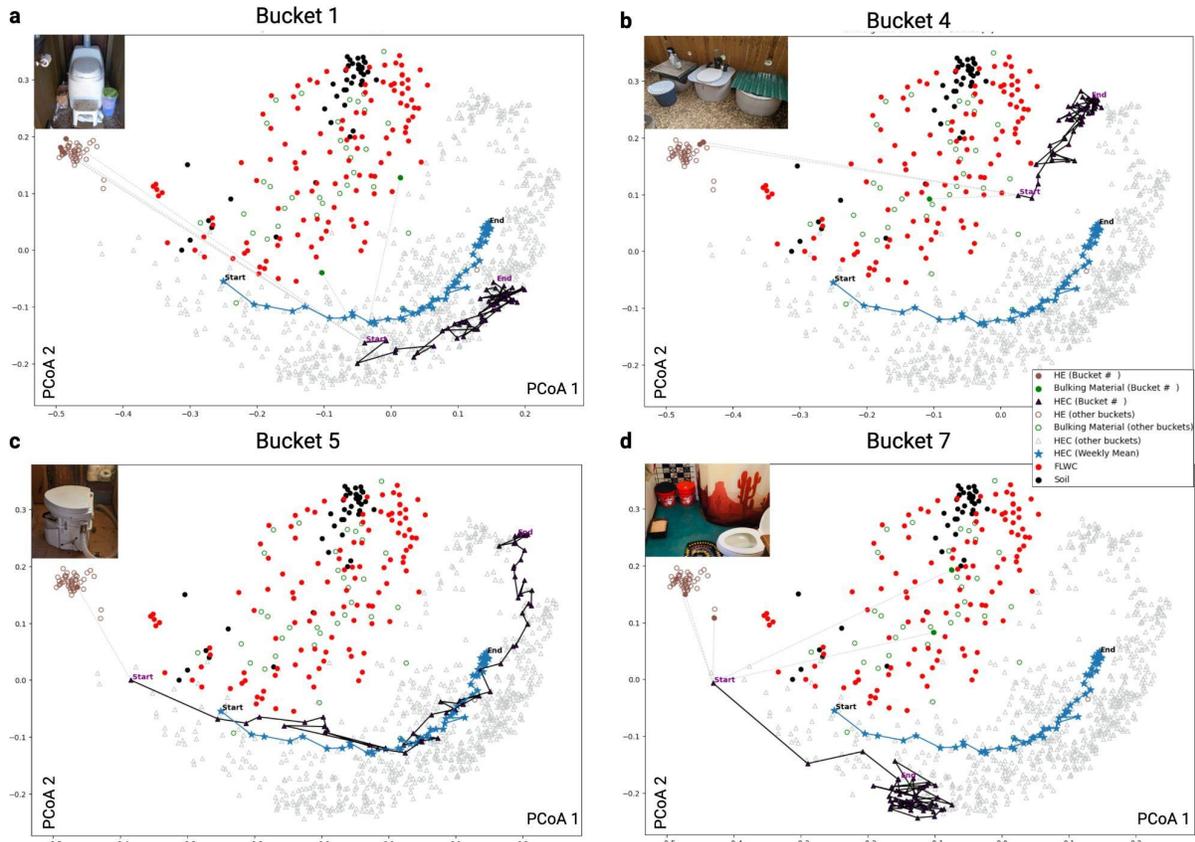

**Figure 1 | Two-dimensional PCoA plots illustrate the trajectory of HEC microbiomes along the gut-to-soil microbiome axis.** Two-dimensional Principal Coordinate Analysis (PCoA) was applied to the Unweighted UniFrac distance matrix computed between all samples including soil and FLWC reference samples. The trajectory through the ordination space for individual buckets is plotted with the statistical mean of ordination results across all 15 buckets (denoted by blue stars and line) at each time point. Averages are not points in the ordination, but rather overlaid on the ordination plot. The average trajectory exhibits a progressive movement away from HE along Axis 1 over the course of one year, with final TPs clustering closer to reference samples. (a) Bucket 1's transition rapidly diverges from HE and its transition appears to stabilize around week 11. (b) Bucket 4's initial transition also occurs rapidly, beginning and ending less similar to fecal material than average. (c) Bucket 5 begins more similar to HE and has the greatest extent of change relative to all buckets. (d) Bucket 7's transition appears stunted, beginning and ending more similar to HE than average. These microbiome composition data do not offer sufficient resolution to provide clear identification of all pathogens or provide insights into microbial activity; future studies will address these aspects. Definitions: *Extent* - Euclidean distance between two data points in the ordination. *Rate* - extent divided by the difference in time between two data points in the ordination. A higher rate is illustrated by consecutive data points spaced farther apart, indicating rapid changes. Conversely, a lower rate is characterized by data points that remain relatively close. *Trajectory* - successional pattern of the microbiomes, illustrated by consecutive data points connected by a black line on the ordination plot for each bucket. Some trajectories span the entire plot, reflecting substantial changes, while others display trajectories that cluster tightly, indicating reduced variability (i.e., stability) in microbiome composition. Plots for all 15 replicates are presented in Extended Data Fig. 3.

B1 exhibits a rapid transition away from HE samples, however, appears to stabilize around TP-11, with minimal further variation (Fig. 1a). The most dominant taxa in B1 by average relative abundance are associated with bioremediation, industrial treatment processes, nitrogen fixation, lignocellulose degradation, and plant growth promotion (Supplemental Text 2).

B4 also exhibits a rapid transition from HE to TP-1, with little change over time relative to the average (Fig. 1b). This CT was unique in that it was an in-ground barrel-based system. The proximity to the soil, prolonged retention of material, and aeration protocol likely introduced microbiomes from soil and finished compost, thereby influencing the composition of TP-1 and the resulting trajectory. This suggests that inoculants derived from well-functioning CTs, soils, or mature compost could accelerate the rate of change away from fecal-like microbiomes. B4 was dominated by taxa associated with compost high in recalcitrant compounds, soils, and agricultural waste (Supplemental Text 2).

B5 was characterized by foul odors (uncommon for properly managed CTs), an initially high pH, and a nearly 100% moisture content which likely induced anaerobic conditions influencing its trajectory. TP-1's microbiome composition initially remained similar to HE, however, it ultimately exhibited the greatest extent of change of all buckets, ending with a microbiome profile similar to reference samples and B4 (Fig. 1c). Dominant taxa were associated with the formation of humic substances, the human gut, and manure (Supplemental Text 2).

The transition of B7 appears stunted, starting and ending with greater similarity to HE than average (Fig. 1d). Weekly samples cluster beginning at TP-4 and do not progress along Axis 1 as observed in other buckets. Dominant taxa were associated with soil, food compost, and polysaccharide degradation, and known to contain ARG (Supplemental Text 2).

We additionally performed qPCR and culturing to track the load of fecal indicator taxa. A significant negative correlation was observed between *E. coli* target copy number and time ($\rho = -0.49$, $p = 2.73e-48$) across all buckets (Fig. 2a and Extended Data Fig. 4), with most dropping below the limit of quantification (LOQ = 304) between TPs 20-35 and remaining below the limit of detection (LOD = 30.4). B5 exhibited relatively high copy numbers which may be due to anaerobic conditions. B15 was the only system where copy numbers were above the LOQ at TP-52, possibly due to regrowth influenced by low temperatures, nutrient availability, or variable environmental factors[7].

There was a significant positive correlation between *C. perfringens* copy number and time ($\rho = 0.28$, $p = 4.96e-16$) across all buckets (Fig. 2b and Extended Data Fig. 4). By TP-52, all buckets except B1 maintained levels above the LOQ. Copy numbers gradually declined in B7 over time but remained above the LOQ throughout the sampling period.

Culturing experiment data demonstrated that all buckets exhibited a decline in *E. coli* with time ($\rho = -0.546$, $p=4.93e-31$), eventually becoming undetectable by TP-25 (Fig. 2c and Extended Data Fig. 5). Detection and quantification of *E. coli* using qPCR after TP-25 likely reflects the method's sensitivity in detecting viable but nonculturable or non-viable *E. coli* DNA[8]. Small-scale composting systems typically do not attain thermophilic temperatures unless they are insulated or externally heated[9], which likely explains why fecal indicator taxa remained detectable. Mesophilic composting can proceed efficiently and reduce pathogens, but longer durations are recommended[10].

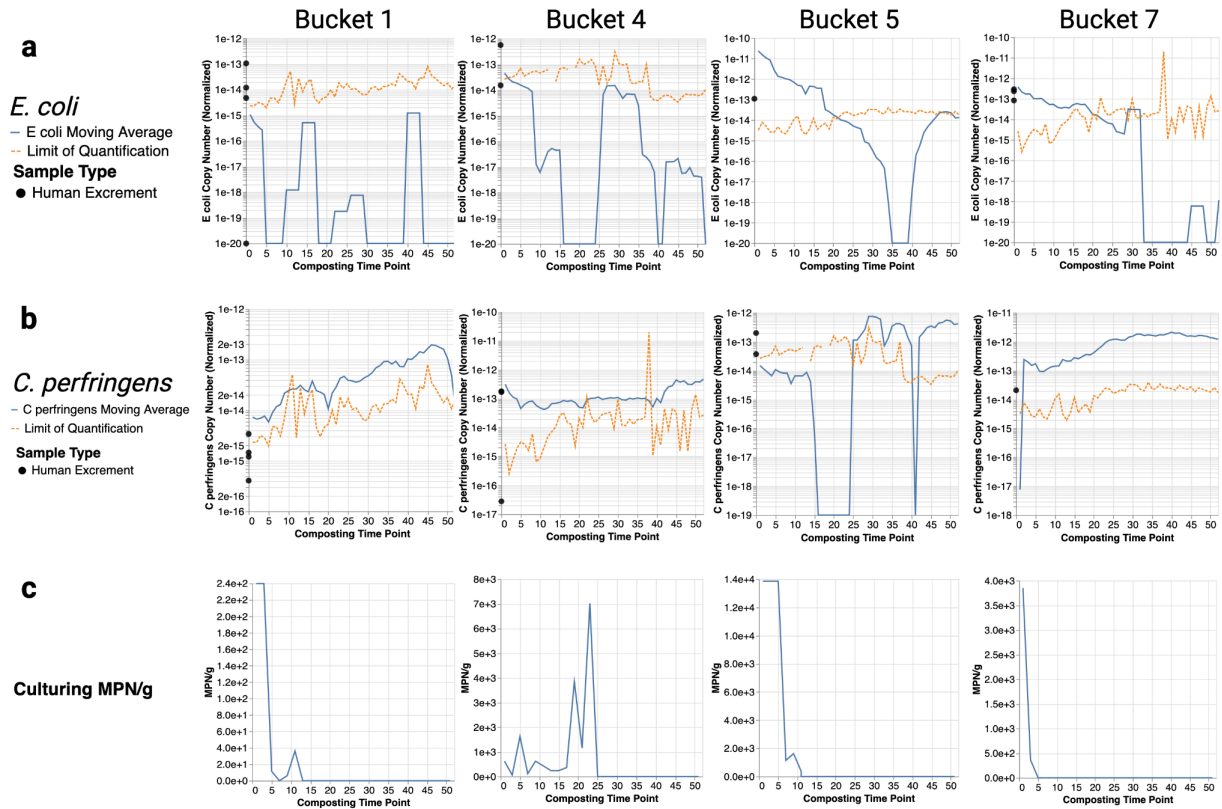

**Fig. 2 | qPCR and culturing data for select buckets.** (a) Copy numbers of the uidA gene found in *E. coli* normalized by 16S rRNA gene copy number data from the BactQuant assay (indicative of total bacterial load). (b) Copy numbers of the *C. perfringens* group normalized by BactQuant. A moving average period of four days was applied to (a) and (b). A pseudo-count, defined as the floor of the log10 of the minimum, minus one, was added to all data points but varied by plot because the minimum nonzero quantification varies by plot. The y-axis scale is not consistent across buckets to highlight individual variability in copy numbers. Significant fluctuations in data are attributed to sampling variability, given that TaqMan PCR analysis was conducted in triplicate. The limit of quantification (LOQ) is 304, and the limit of detection (LOD) is 30.4 and are presented here also normalized by BactQuant (orange lines). Raw results of the BactQuant assay are presented in Extended Data Fig. 6. Normalized data for all buckets can be found in Extended Data Fig. 4 and non-normalized data for all buckets can be found in Fig. S8. (c) Biweekly samples (odd TPs) were cultured in EC broth with MUG for the detection of *E. coli*. Data is presented linearly and the y-axis is not consistent across buckets to highlight the variability in the composting systems. Culturing data for all buckets can be found in Extended Data Fig. 5.

Our primary hypothesis, that the microbiome of 1 year one-year-old HEC would resemble that of soil and FLWC, was supported by the data (Fig. 1 and Fig. S6). Our secondary hypothesis, that fecal indicators *E. coli* and *C. perfringens* would be undetectable after one year of mesophilic composting, was supported only by *E. coli* culturing data but not the qPCR results for *E. coli* and *C. perfringens*.

This investigation identified significant shifts in microbial communities around 25 weeks, with reductions in fecal-associated taxa and increases in environmental taxa linked to nutrient cycling and organic matter decomposition, indicating effective composting[2,3] (Extended Data Fig. 7 and Supplemental Text 2). Reduced rates of microbial succession by TP-52 may result from stabilization due to uniform physicochemical conditions, reduced nutrient availability, or microbial interactions[11].

The unique trajectories of B4 and B7 provide insights into CT design and management. B4 rapidly developed a microbiome similar to reference samples, suggesting its design and maintenance protocol may optimize composting efficiency. In contrast, B7 showed stalled microbial succession potentially due to improper CT use or suboptimal ecological niches. To facilitate safe HEC, we need to understand why a transition might stall, whether that indicates safety concerns, and if so, how to reboot its transition. Contrasting the microbiome or other features of B7 with B4—or B6, (B6 and B7 were provided by a cohabiting couple), which temporarily stalled at a similar TP but eventually continued its transition along PCoA axis 1 (Extended Data Fig. 7b and Supplemental Text 2)—might reveal methods for remediating underperforming systems such as inoculation.

Our research identified microbial succession patterns that could serve as biomarkers for HEC completion and safety. Key taxa such as *Rhodanobacter, Rhodococcus, Arachidicoccus, Acinetobacter*, *Pseudomonas, Hyphomicrobium,* and members of the Rhizobiaceae family demonstrate essential roles in nutrient cycling and could serve as bioindicators to determine if microbial communities will stabilize or shift (Extended Data Fig. 8 and Supplemental Text 2). Furthermore, the high diversity of taxa observed suggests that HEC microbiomes may prove fruitful for bioprospecting for enzymes or microbial consortia useful in decomposing materials potentially harmful to human and/or environmental health such as pharmaceuticals or PFAS.

*E. coli* was undetectable by 25 weeks in culturing experiments, but its slow decline via qPCR suggests potential risks and methodological challenges. Conversely, *C. perfringens* levels increased, suggesting mesophilic systems may support the persistence or regrowth of endospore forming organisms consistent with other studies[2,3,7]. While not all *E. coli* and *Clostridium* species are pathogenic, these data reinforce careful HEC risk management, particularly while transferring material into thermophilic piles for further decomposition and pathogen reduction[3]. Guidance from resources like *The Humanure Handbook*[12] can assist CT users worldwide to safely manage HEC. Many CT manufacturers currently provide inadequate, inconsistent, or ambiguous HEC management instructions, creating potential for greenwashing and improper use[13].

There is an urgent need to re-evaluate broad applicability of HEC amid rising challenges from climate change and population growth. This study comprehensively assesses microbial succession during mesophilic composting of HE, addressing a key gap in understanding the gut-to-soil microbiome transition in HEC. This study also provides the highest resolution composting microbiome data to date, establishing a baseline for HEC optimization and thermophilic composting studies while serving as a resource for bioprospecting organisms relevant to upcycling waste.

# Methods

We recruited 13 individuals (Northern Arizona University IRB protocol: 1773199-3) who were existing CT users in Arizona and Colorado, USA, to use their CT for three weeks (the typical time it takes to fill a personal CT system) and provide the resulting material (a combination of their HE and bulking material (BM) - a carbon-rich material such as sawdust) to us for weekly sampling over the course of one year (Extended Data Fig. 1 and Supplementary Text 1). By having participants use their existing systems, we aimed to increase the number of participants by not requiring them to deviate from their normal habits. This however introduced variability across replicates, resulting from different systems, bulking materials, and usage habits. In a baseline study such as this, we consider this a tolerable source of variation as it allows us to evaluate systems across real-world usage scenarios. Variables that differed across systems and potentially influenced the composting microbiomes included the geographic and physical location of the CT; participant's diets, defecation patterns, and volumes; styles of CTs; and the volume and type of BM used.

Upon collection of material from CTs following usage, we transferred the material to modified 19-liter buckets that included a sampling access port and ventilation tube to allow air flow in the buckets. Buckets were stored in a climate controlled greenhouse and were rolled weekly in a consistent manner to further aerate and homogenize the contents. Just after rolling, 2-gram samples were collected through the access port and physicochemical properties of HEC (including temperature, moisture, and pH) and qualitative variables were recorded weekly. In addition to the weekly sampling, participants provided weekly HE (i.e., fecal) samples during their three-week collection period, a sample of the inside of their CT before they began usage, a sample of their BM, and a sample of the area surrounding the CT (by swabbing the ground with a sterile swab).

Our 13 participants provided a total of 15 HEC biological replicates. Two individuals participated three times each, collectively contributing six buckets and two participants contributed to bucket 10 (i.e., B10). All other individuals participated one time. Two cohabiting couples participated in the study (contributing B2 and B3, and B6 and B7).

HEC began at time point 1 (TP-1) for each bucket. Because participants began their collection at different times, time points represent time since the start of HEC on a per bucket basis and do not represent consistent calendar dates across buckets.

Analyses of samples included: microbiome profiling by 16S rRNA gene sequencing[14]; qPCR assays for *E.coli* and *C.perf*[15,16]; a pan-enterovirus assay[17] on pooled HE samples only; a BactQuant assay[18]; and culturing *E. coli* in EC broth with MUG[19] (Supplementary Text 2).

# Data Availability

Raw and processed data for this study is available in Zenodo under DOI 10.5281/zenodo.13887457.

# Code Availability

Custom software was developed to create Figures 1 and 2. This code is available in a QIIME 2 "Single Use Plugin" at https://github.com/caporaso-lab/gut-to-soil-manuscript-figures, and in Zenodo under DOI 10.5281/zenodo.13887457.

# Acknowledgements

The authors wish to thank all research participants for their HE compost samples.

# Inclusion & Ethics Statement

This study was conducted with a commitment to ethical principles and inclusivity. Research design, data collection, and analysis were guided by considerations for equity and respect for diverse perspectives. The authors ensured that contributions were acknowledged equitably, fostering an inclusive environment for all team members. We adhered to ethical standards set by our institution's human subjects research review board (IRB approval ID 1773199-3) and followed guidelines to ensure responsible handling and dissemination of data.

Data was anonymized to protect participants' privacy, and results were communicated transparently to avoid misinterpretation or bias. All collaborators were provided opportunities for input throughout the research and publication process. Our work upholds Nature's ethics guidelines and emphasizes the importance of inclusivity and integrity in scientific advancement.

# Author Contributions

J.M., J.G.C., D.J., N.S. and C.W. designed the project. J.M., C.H., A.O., G.A., D.B., E.B., K.C. H.C., J.C., T.D., A.D., D.E., M.F., L.G., J.H., C.H., P.I., K.K., V.L, C.O., V.Q., D.R., A.R., M.R., J.S., S.S., N.S., S.T., K.V.O., A.P., S.W., L.W., C.W. and J.Y. collected and processed data and/or constructed the figures and tables. J.M., C.H., A.O., G.A., D.B., E.B., K.C. H.C., J.C., T.D., A.D., D.E., M.F., L.G., J.H., C.H., P.I., D.J., K.K., P.K., V.L, C.O., V.Q., D.R., A.R., M.R., J.S., S.S., N.S., S.T., K.V.O., A.P., S.W., L.W., C.W., J.Y. and J.G.C. contributed to writing, reviewing and editing the manuscript.

# Competing Interest Declaration

**The authors declare no competing interests.**

# Additional Information

**Extended Data** is available for this paper at:

**Supplementary information** The online version contains supplementary material available at:
**Correspondence and requests for materials** should be addressed to Greg Caporaso or Jeff Meilander

**Peer review information**

**Reprints and permissions information** is available at www.nature.com/reprints.

# EXTENDED DATA

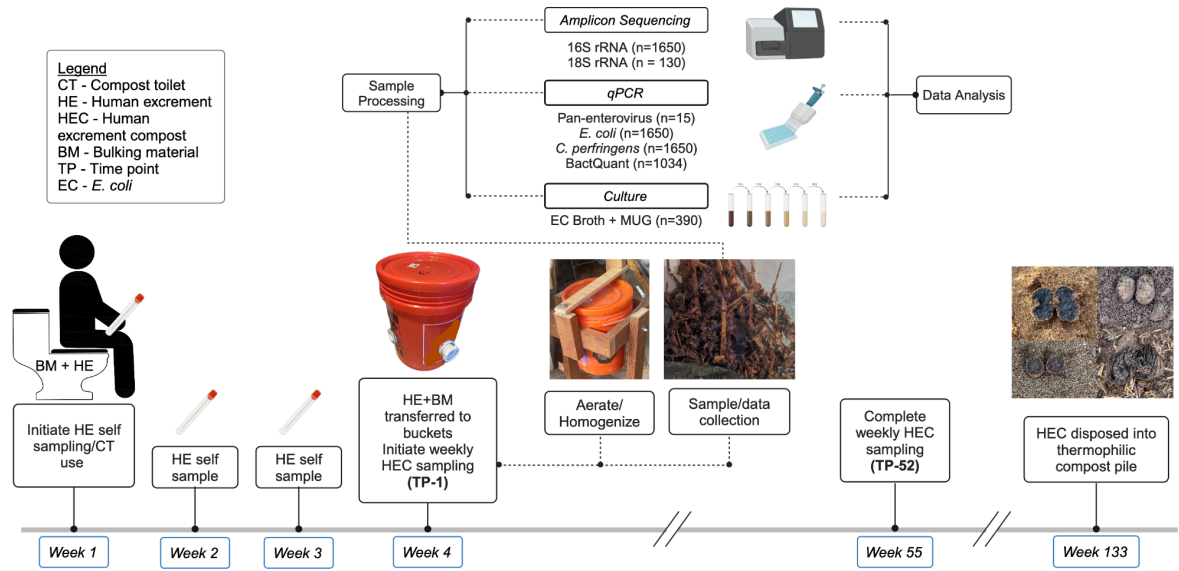

**Extended Data Fig. 1 | Experimental design definitions and workflow.** *Participants* - individuals (aged 33-73 years) contributing to the study (11 total); completed a self-reporting questionnaire. *Human excrement (HE)* - fresh samples from participants collected on sterile cotton swabs. *Composting toilets (CT)* - alternative toilet systems used to collect HE. Urine diversion was required and all toilets were batch style CTs. *Bulking material (BM)* - carbon rich material to balance carbon to nitrogen ratio (C:N) and reduce odors. *Human excrement compost (HEC)* - mixture of HE and BM. *Bucket* - a 19 L bucket modified with a ventilation tube and access port in which the HEC was collected, stored, and sampled from during the sampling period (52 weeks). Individual buckets are biological replicates, referenced as B# (i.e. B1, B2). Buckets were stored in a crate within a temperature controlled greenhouse between 13–16°C. *Composting* - human-directed process of optimizing microbiomes in the aerobic biodegradation of organic wastes at thermophilic temperatures, resulting in a safe, mature, stable, and nutrient-rich soil amendment. Thermophilic temperatures were not targeted nor achieved in this experiment. *Food and landscape waste compost (FLWC)* - reference samples including pre- and post-consumer food waste co-composted with landscaping waste such as leaves, pine needles, branches, and straw collected from university dining halls, campus, and local residences. *Soil* - Earth Microbiome Project reference samples downloaded from publicly available data on Qiita. *Time point (TP-#)* - indicates the HEC point of sampling after HEC was collected from CTs. **Workflow:** Compost toilet usage commenced at week 1, with weekly sampling initiated at week 4, referred to as TP-1. Aeration and sample collection occurred weekly until TP-52, with samples processed in batches. HEC remained inside the buckets until week 133, at which point the material was transferred into an actively managed thermophilic compost pile and allowed to compost for another year. Created in BioRender.

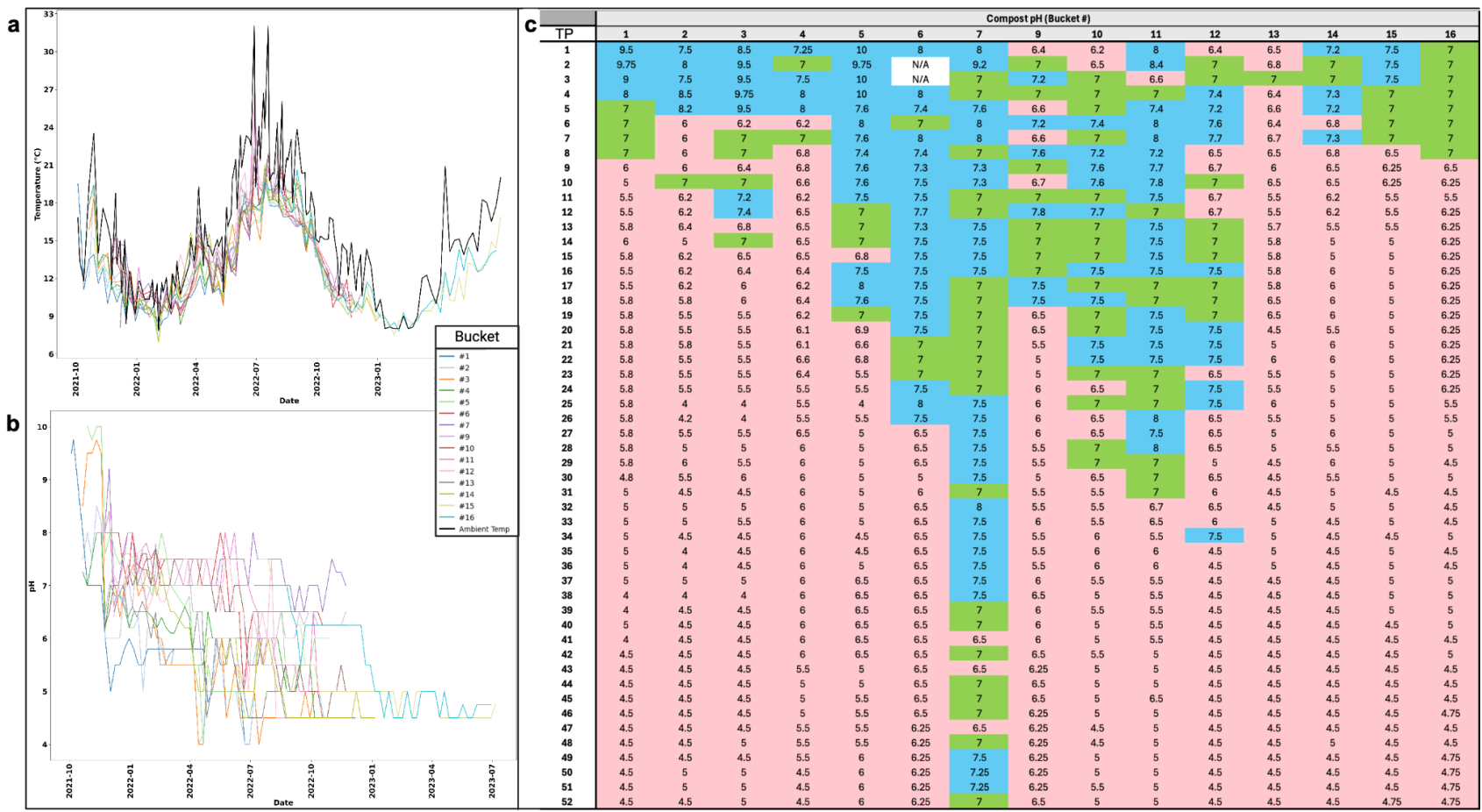

**Extended Data Fig. 2 | Temperature and pH of HEC.** (a) The temperatures of human excrement compost (HEC) for individual buckets (represented by colored lines) aligned with ambient temperatures (black line) in the greenhouse. (b) The pH of HEC for individual buckets (represented by colored lines) fluctuated weekly but exhibited an overall decrease. (c) Although oscillations in pH occurred over time the pH quickly became acidic and trended downward for all Buckets, except S7, ending with a neutral pH. This likely indicates unfinished composting. Blue indicates basic, green indicates neutral, and red indicates acidic pH levels. Created in BioRender.

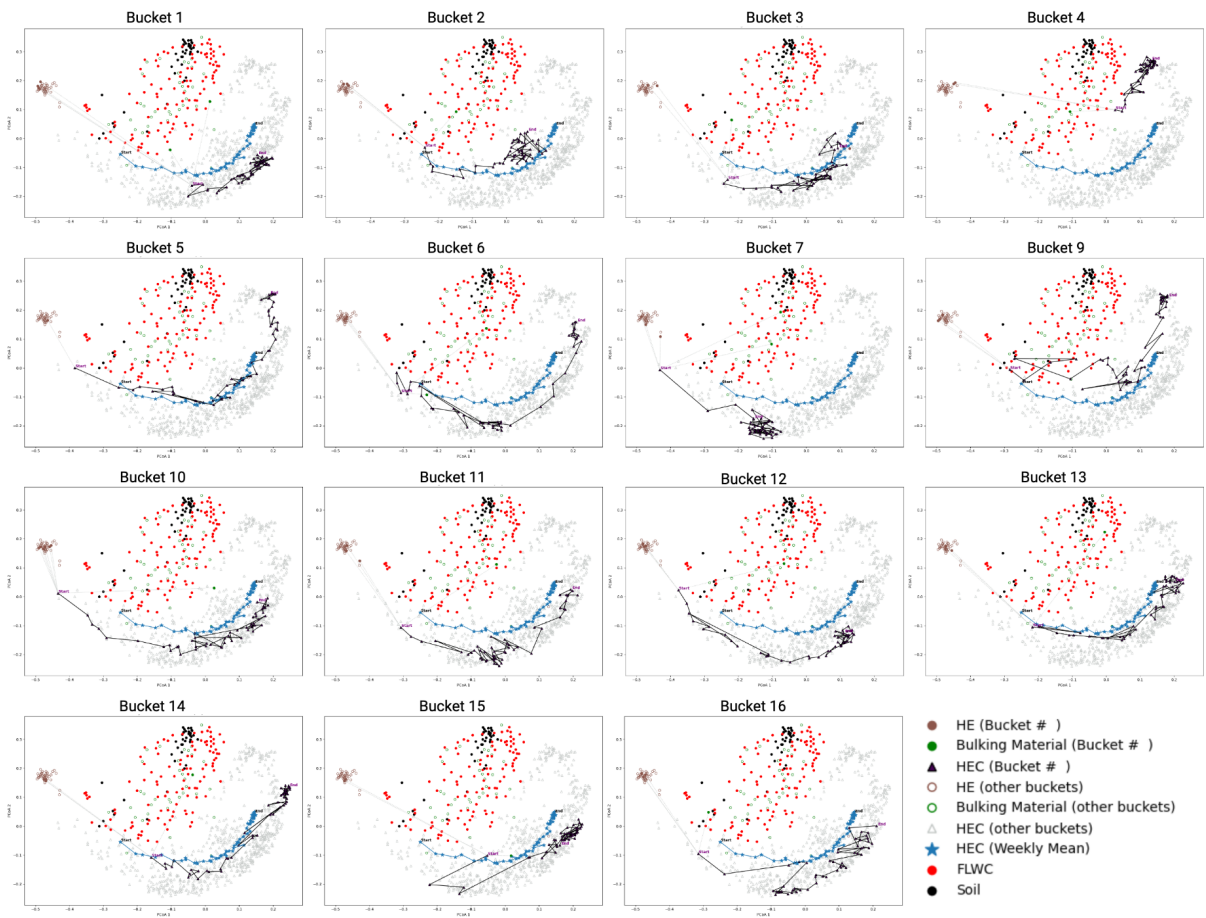

**Extended Data Fig. 3 | Ordination results of unweighted UniFrac distances for all buckets.** Principal coordinates plots of Unweighted UniFrac distances between samples from each bucket were plotted against the statistical mean of ordination results for all 15 buckets (denoted by blue stars and line - "average") at each timepoint illustrating variability in the transition in microbiome composition of mesophilic human excrement compost (HEC) over 52 weeks. A black line connects consecutive TPs from TP-1 ("Start") to TP-52 ("End") illustrating the microbial succession over time. The HE (brown) and BM (green) samples from each bucket are connected to TP-1 by a dotted line to illustrate the inputs of the system. Red and black dots represent the reference samples and form a diffuse cluster (along with BM) in the center of the plot. Created in BioRender.

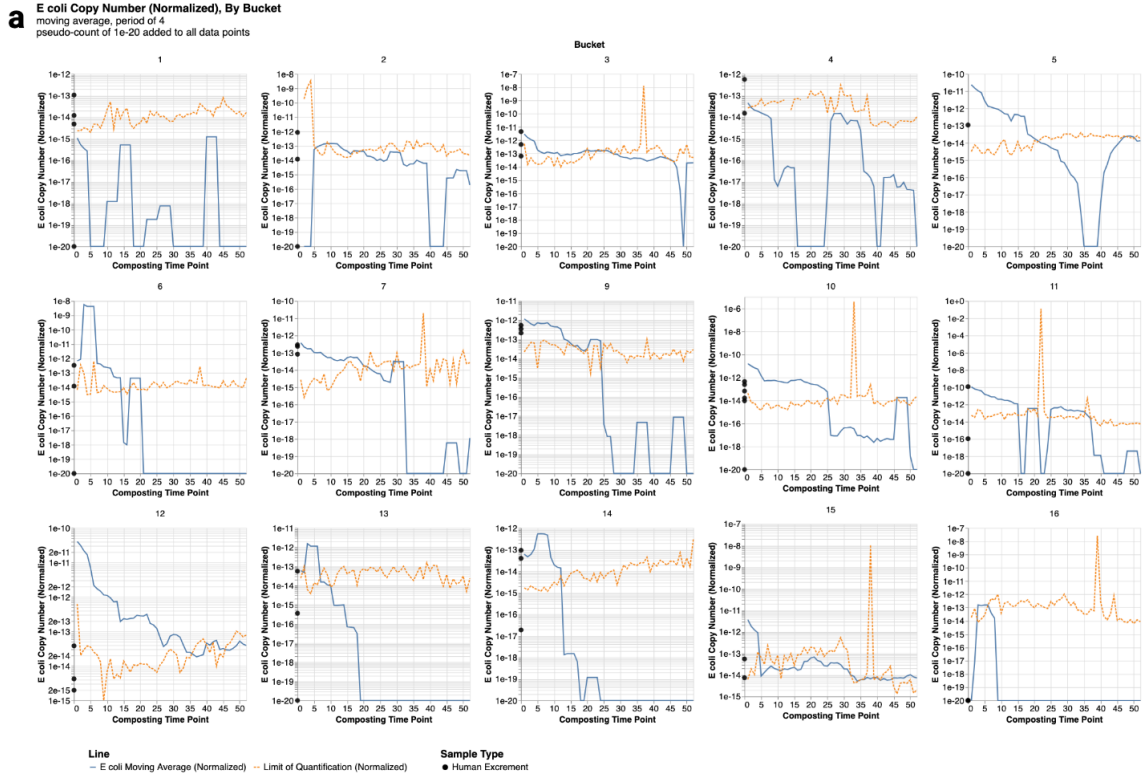
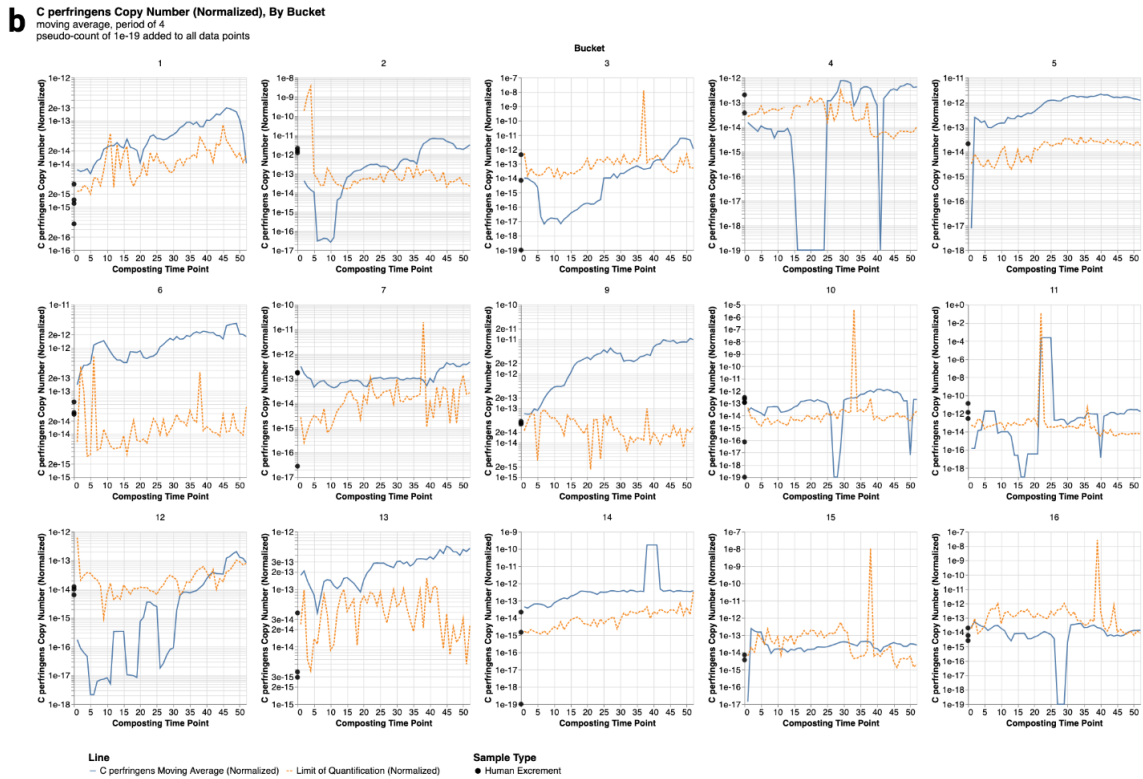

**Extended Data Fig. 4 | Normalized qPCR results for all buckets.** (a) *E. coli* and (b) *C. perfringens*. Created in BioRender.

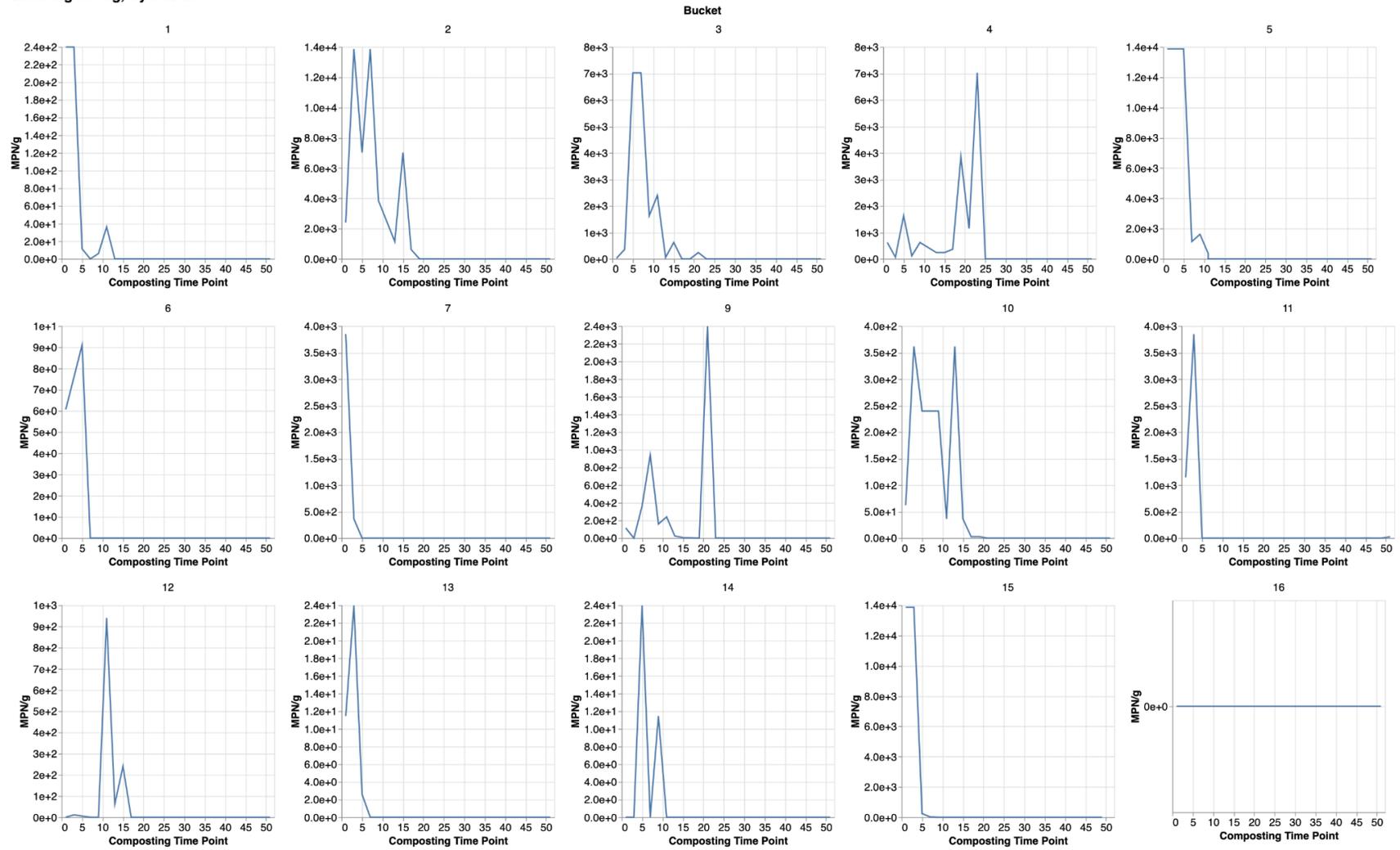

**Extended Data Fig. 5 | Culturing results for all buckets.**

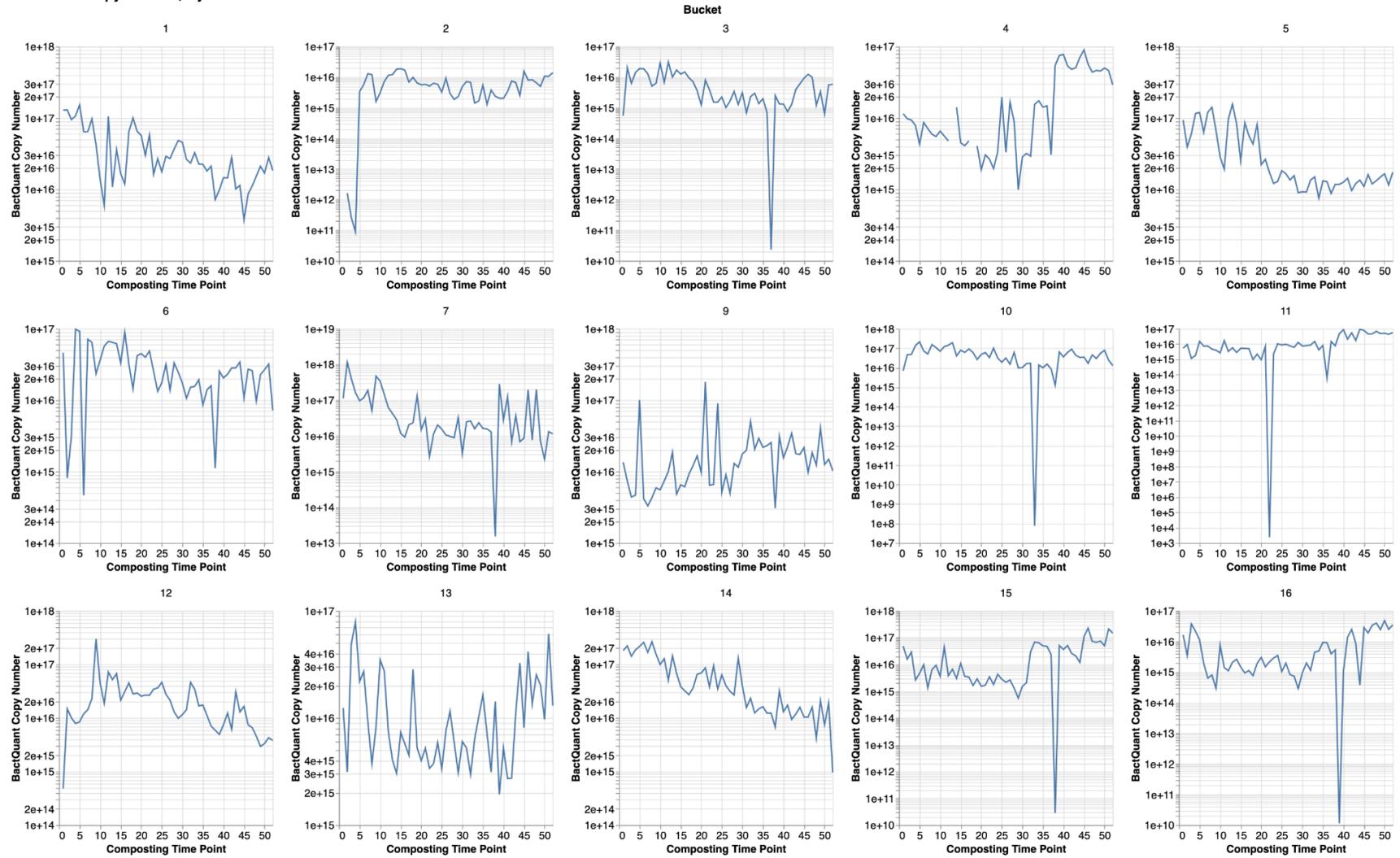

**Extended Data Fig. 6 | BactQuant assay results.**

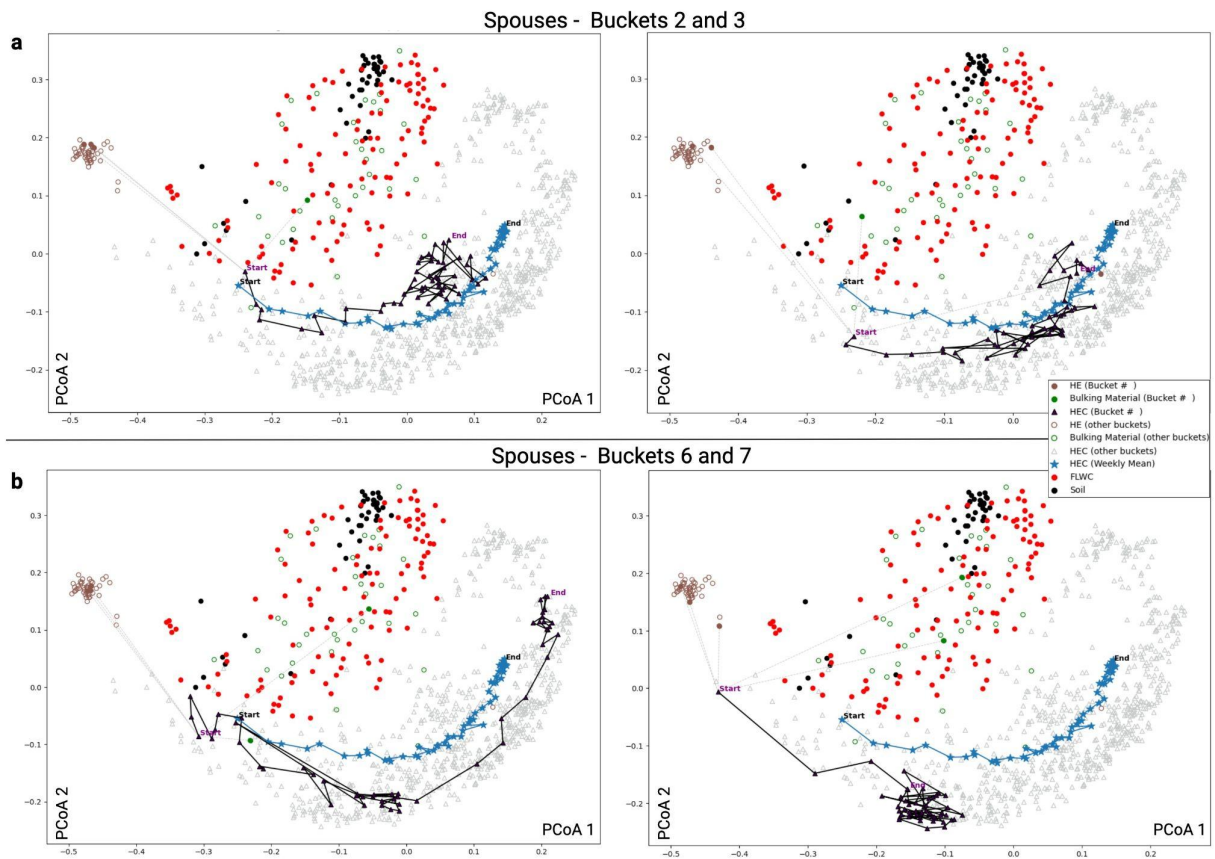

**Extended Data Fig. 7 | Ordination results of unweighted UniFrac distances for spouses.** (a) Buckets 2 (left) and 3 (right), created by spouses, follow a similar trajectory from start to finish. In contrast, (b) buckets 6 (left) and 7 (right), also created by spouses, initially show similar stalling patterns, but bucket 6 undergoes a rapid shift in trajectory. This highlights variability in microbiome succession patterns between spouses. Created in BioRender.

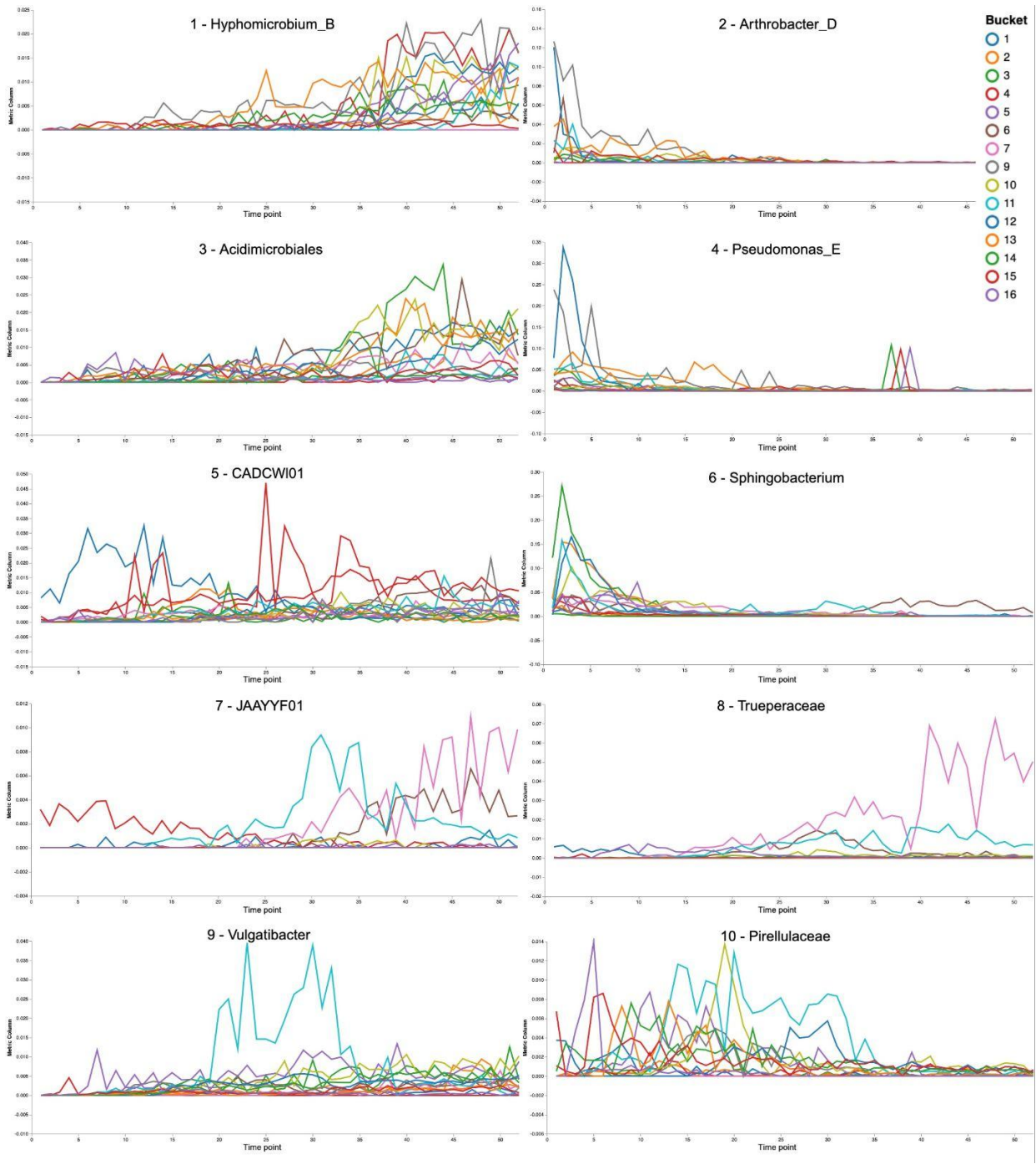

**Extended Data Fig. 8 | Feature volatility plots displaying the top 10 features. Created in BioRender.**

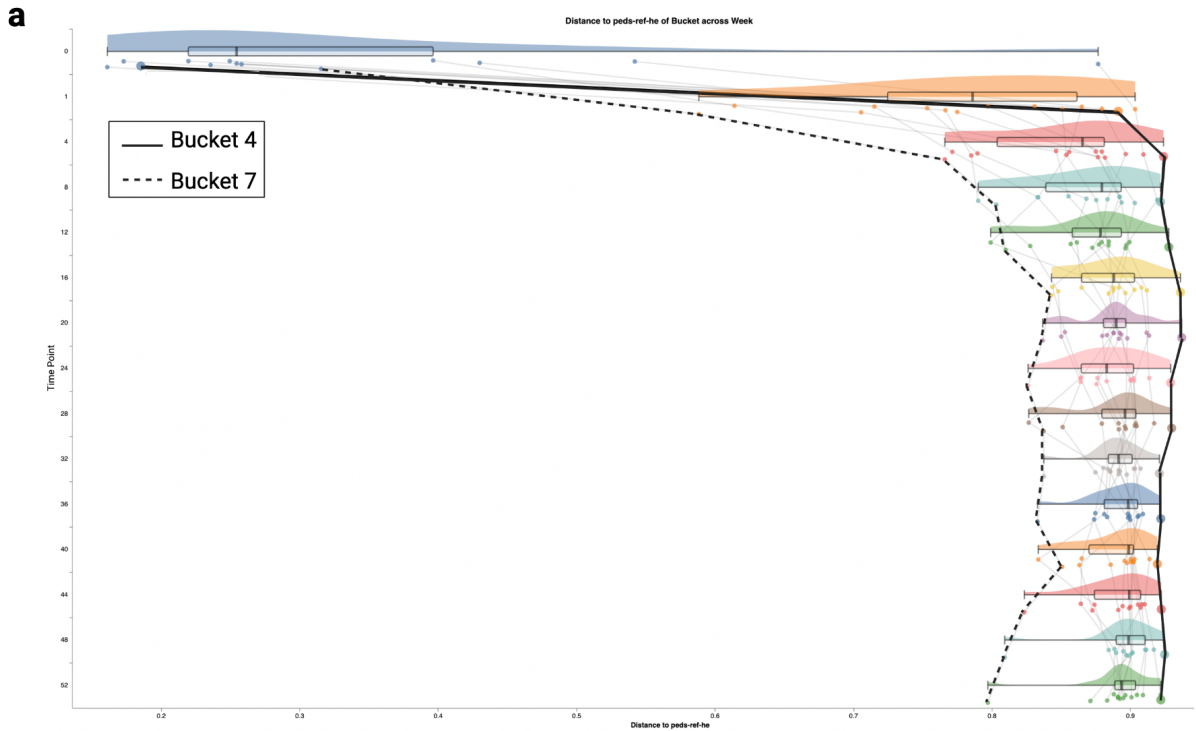

**Extended Data Fig. 9 | q2-FMT results.** (a) q2-FMT raincloud plots showing the distribution of buckets' measure of distance to peds-ref-he across time points. Kernel density estimation performed using a bandwidth calculated by Scott's method. Boxplots show the min and max of the data (whiskers) as well as the first, second (median), and third quartiles (box). Points and connecting lines represent individual buckets with a consistent jitter added across groups such that slopes across adjacent groups are visually comparable between buckets. The x-axis implies a shift only due to time, however, the observable shift is also due to a difference in sample types; i.e. the BM in the CT initially and the HE. The blue cloud at TP-0 encompasses all data points for all buckets, comparing the Unweighted UniFrac distances between one HE sample and other HE samples for each bucket, thereby establishing a baseline for comparing HEC to HE. The first orange cloud represents the change in Unweighted UniFrac distances between TP-0 (HE) and TP-1 (HEC). At TP-1, the Unweighted UniFrac distances of each bucket's HEC samples quickly become more dissimilar to their respective HE, shifting the cloud to the right on the x-axis. Subsequent TPs become increasingly dissimilar over time (some reaching 93%) and display reduced variation across buckets, as indicated by the narrowing of the box plots. Buckets 4 and 7, outliers compared to all buckets are highlighted by a solid black and dotted black lines respectively. (b) Wilcoxon Signed Rank tests between groups (time point vs time point), with two-sided, asymptotic p-value calculations and Benjamini–Hochberg correction for multiple comparisons (q-value). Created in BioRender.

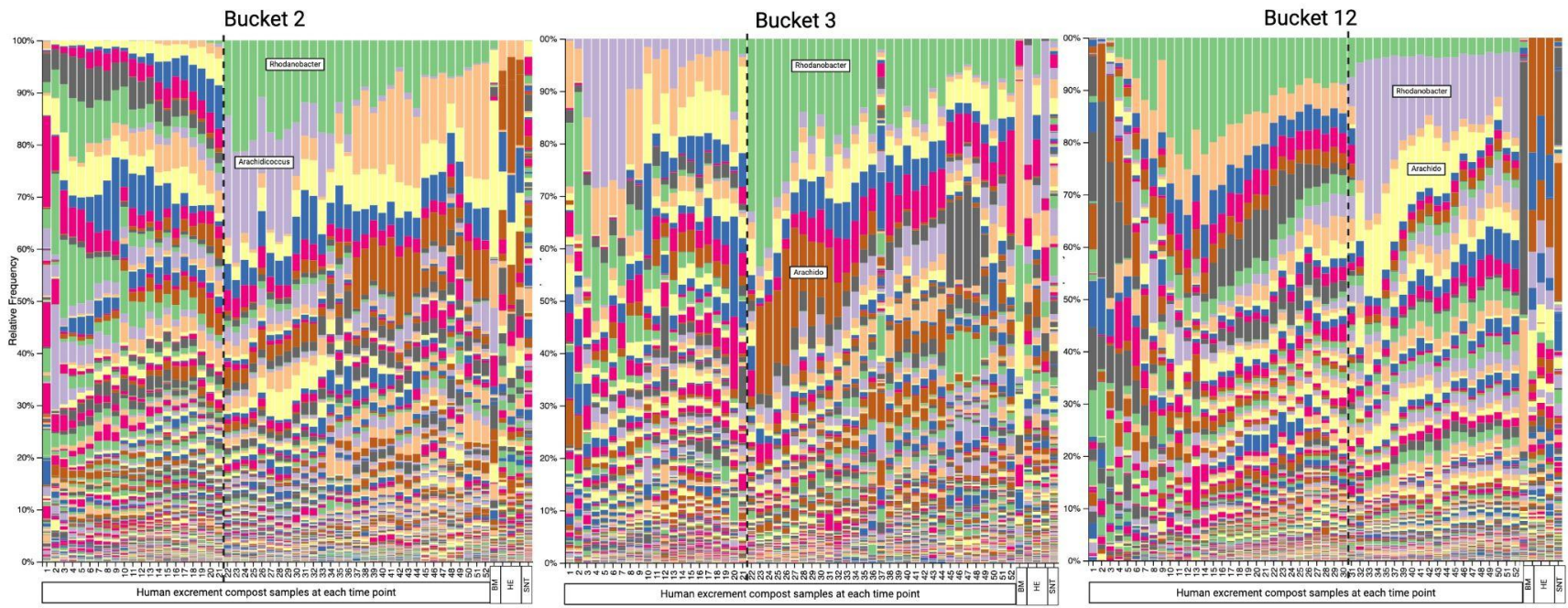

**Extended Data Fig. 10 | Genus-level taxa bar plots of Buckets 2, 3, and 12.** The dotted lines mark time points with a notable increase in the relative abundance of *Rhodanobacter* (green B2 and B3 and purple B12) and *Arachidicoccus* (purple B2, brown B3, yellow B12). This pattern was not observed in other buckets, even those created by the same participants. BM-bulking material; HE-human excrement; SNT-soil near toilet. Created in BioRender.

# Supplementary Text 1: Human Excrement Compost Workflow

## 1.1 Participants and compost toilet systems.

Participants submitted a health data survey (Table S1) and a description of their compost toilet system (Table S2).

## 1.2 Bucket design and rotation

19-liter buckets were modified for this experiment (Fig. S1a). To facilitate aeration a 3.81 cm PVC tube was inserted 13.97 cm from the bottom of the bucket. Ten holes (five on each side) measuring 9.5 mm were drilled through the tube. The tubes were secured in place (holes oriented horizontally) by 6.35 mm PVC couplers affixed to the tube's exterior using Oatey's Red Hot Blue Glue. Silicone caulk sealed any gaps between the bucket and the tube. Finally, window screen was taped over the tube opening to prevent HEC or insects from leaving the tube.

To facilitate sampling, an access port was added perpendicular to the aeration tube, situated 10.16 cm above the bucket's bottom, using a 5.08 cm threaded reducing bushing with a 3.81 cm threaded plug. The placement allowed access to the HEC as it compacted during composting.

HEC was removed from CTs after three weeks of use and transferred to the Bucket. The Buckets were stored in a storage crate located inside a climate controlled (12.5°C) greenhouse in the South Campus Greenhouse Complex at Northern Arizona University.

A 1.73 $m^3$ storage crate for the buckets was fabricated from repurposed pallets and plywood, elevated 9 cm above ground. The interior was lined with a 10 mm thick tarp to mitigate the risk of potential spills. The crate had three fixed sides, while the fourth side, functioning as the door, was exclusively connected to the crate floor via three 7.6 cm hinges. Lowering the door created a workspace for sample collection. The buckets were systematically arranged in stacks of three, advancing through 20 locations within the crate each week (Fig. S1b). Plastic totes stored sampling equipment and were kept inside the crate between sampling periods.

## 1.3 Reference Samples

### 1.3.1 Food and Landscape Waste Compost (FLWC)

In a previous, unpublished study, FLWC samples were collected from a composting facility located on the Northern Arizona University campus, six small agricultural farms, and two backyard gardeners in Flagstaff, AZ. Samples were collected in triplicate 14 cm below the surface of each compost pile, except at the university compost facility where, due to the large size of the piles, samples were collected in triplicate 45.7 cm below the surface. A 30.5 cm hand trowel, cleaned and sanitized between each collection, was entirely inserted into the pile. One full trowel was removed, the material placed into a 3.8 L sealable plastic bag, and chilled on blue ice for up to two hours before being transferred to a -20°C freezer. A total of 154 samples were collected from piles aging between 0 months (freshly constructed) and 2 years.

### 1.3.2 Soil

Sequences (150 bp, single end reads) associated with 41 Earth Microbiome Project (EMP) soil samples[1] from the "EMP500" collection were downloaded from publicly available data on Qiita (https://qiita.ucsd.edu; study: 13114) to be used as reference samples[2].

## 1.4 Collecting HEC from Participants

- Upon arrival at the participant's toilet location, put on PPE (mask and gloves)
- Prepare the transfer site
  - Lay down a small tarp or 6 mm plastic
  - Set the empty 19 L modified bucket in the center of the plastic with the lid removed
  - Spray the inside with 70% ethanol and allow to air dry
  - Sanitize the external surface of the composting toilet
- Transfer the material
  - Open the composting toilet carefully to ensure there is no spillage
  - Empty the composting toilet collection chamber into the 19 L bucket
  - Close the composting toilet carefully
  - Sanitize the external surface of the composting toilet
  - Replace the lid on the 19 L bucket making sure the lid is tightly sealed
  - Sweep up any spilled material into a small pile
  - Sanitize the external surface of the bucket
  - Place the 19 L bucket into a large black trash bag and set to the side of the plastic on the ground
  - Carefully roll up tarp or plastic to keep any spills in the center and place into a separate garbage bag
  - Triple bag the 19 L bucket and finally place it into a large Tupperware tote for transport
  - Place the garbage bag with the plastic into the tote for transport
  - Remove gloves and mask and place into the garbage bag
  - Wash hands and use hand sanitizer
- Deliver the material
  - Deliver the 19 L bucket to the sampling site in the Greenhouse Complex on South campus of NAU
  - Put on mask and gloves
  - Remove the bucket from the three garbage bags
  - Sanitize the external surface of the bucket
  - Place the bucket in the respective location in the sampling crate
  - Label the bucket with the date, location code, and a biohazard sticker
  - Place the three garbage bags, gloves, and mask into the garbage

## 1.5 Sampling

A selected bucket was removed from its designated location in the storage crate, weighed and ambient temperatures, and external observations were recorded. External views of each bucket and through the ventilation tube were photographed. Subsequently, the bucket was placed into a bucket roller constructed from recycled lumber for aeration and homogenization (Fig. S1c). The rolling pattern comprised five rotations along the x-axis, five counterclockwise rotations along the z-axis (with the bucket on its side), five rotations along the y-axis, and five clockwise rotations along the z-axis (with the bucket on its side).

After homogenization, buckets were returned to the sampling platform. The access port, sampling platform, and utensils (spoon or tongs) were first sanitized with 70% ethanol followed by 10% bleach. Sealable sample collection baggies (sandwich) were labeled with a unique, eight character identifier, date, and weighed. The access port was opened, and a photo was taken of the contents. Compost temperature and moisture were measured using a digital thermometer (Comluck) and a VH400 soil moisture sensor (Vegetronix) attached to a multimeter (Klein Tools). Moisture data was excluded from analysis due to the variability of probe readings as a result of the heterogeneous HEC. Some recordings were taken from dense, compact areas (e.g. HE mass and high moisture), while others were collected from regions with large air pockets and less moisture, resulting in inconsistent measurements. To minimize rapid depletion of HEC and potential disruptions to microbial succession by removing larger amounts of sample, the gravimetric method[3] for assessing compost moisture content was avoided. Qualitative data including the presence or absence of organisms, compost contents, odor, perceived moisture, and color were also recorded.

Weekly sampling occurred on Tuesday (buckets 1-5 and 14) and Thursday mornings (buckets 6-13 and 16) between 8 am and 12 pm. Fifteen centimeter tongs or 20 cm spoons were used to collect 2.0 gram HEC samples depending on BM used. The sample was extracted, placed into the plastic bag, immediately photographed, and assessed for odor, color, perceived moisture, sample contents, and the presence or absence of organisms. An additional 0.2 grams of the HEC was mixed with 0.125 mL of distilled water, and Asurion 0.5 pH paper was used to measure pH.

After the access port was closed, the outside of the bucket was disinfected and returned to its original location in the crate. The sampling area was cleaned and disinfected before sampling the next bucket.

Post sampling, the entire workspace was cleaned and disinfected and only on Thursdays, bucket advanced one location in the crate according to the bucket rotation guide (Fig. S1b), exposing them to different microclimates and sun exposure throughout the year.

The lids remained sealed except during periodic interventions. For high moisture or foul-smelling buckets (B5, B6, and B7), 100 grams of pine shavings were added post-sampling. Conversely, buckets showing low moisture content (B2, B3, B4, and B15) received 100 mL of distilled water through the access port post-sample extraction. After remediation, buckets underwent an additional rolling procedure as previously outlined.

Bags containing samples were transported to the Genetics Core Lab Facility in the Science Lab Facility Building at Northern Arizona University and securely stored in a -20°C freezer.

### 1.5.1 Weekly HEC Sample Collection Instructions for Researchers

Be sure that appropriate PPE is worn at all times during sampling.
"Ethanol" = 70% ethanol
"Bleach" = 10% bleach
"Sanitize" = spray with 70% ethanol, wipe, and air dry followed by spray with 10% bleach, wipe, and air dry.

Workspace Set-Up
- Put on all appropriate PPE
- Open the latches and drop the wooden door of the crate.
- Open the lid of the small white trash can and place it to the right of the sampling workspace of the crate.
- Pull the black storage bin on the left side of the crate forward and remove the yellow top.
- Remove the top black shelf with sampling tools as well as paper towels, ethanol, bleach, baggies, gloves, and distilled water.
- Replace yellow top and sanitize. Place the sampling spoons, moisture probe, and thermometer on the yellow lid and sanitize each.
- Turn the dial on the orange multimeter 3 clicks to the left (arrow on 20) and place the red alligator clip on the positive terminal of the 9V battery.
- Set the large scale to the right of the sampling space.
- Set up the computer and open the spreadsheet outside of the workspace.
- Before you begin, highlight the sample row titled "HEC" in yellow for each of that day's buckets to reduce errors when entering data.
- For each bucket, weigh a sandwich bag and label it with; the unique sample ID, date, bucket number, and weight of the bag (i.e. 2.0 g).
- Set up the bucket roller behind the sampling workspace.

External Observations
- Put on gloves.
- Place the small pieces of paper labeled 1-7 on the top of the buckets according to the location of that stack. This will help when accessing buckets and ensuring the stacks of buckets are returned to their correct locations.
- Refer to the (Crate Location) column in the spreadsheet and the laminated "Bucket Rotation Diagram" in the crate in order to retrieve the correct bucket.
- Press ON/TARE on the large scale and set the units to Kg. Place the bucket on the scale and record the weight in the (Wt Bucket Compost Kg) column in the spreadsheet.
- In the (Wt Compost Kg) column, be sure the equation is subtracting the bucket weight from the (Wt Bucket Compost Kg). This can be found in the green cell at the top of this column for each bucket.

Bucket Roller
- Place the bucket in the top compartment of the roller with the access port in the back-right corner (Fig. S1c).
- Do a half-turn and give a few small "love-taps" on the bottom to loosen the material.
- Roll bucket 5 times.
- Remove the bucket and place it horizontally on the casters below with the lid facing outward.
- Roll bucket 5 times away from the researcher.
- Return the bucket to the top compartment of the roller, placing the access port in the back-left corner.
- Roll bucket 5 times.
- Remove the bucket and place it horizontally on the casters below with the lid facing outward.
- Roll bucket 5 times toward the researcher.
- Return the bucket to the workspace area and clean the surrounding area.
- Sweep up any material and place it into the garbage bag.

Sample Collection Set-Up

- Place thermometers on top of the bucket and on the floor, wait one minute and record the temps in the (Temp Top Bucket C) and (Temp Bottom Bucket C) columns respectively.
- Make any observations of the outside of the bucket (flies, odor, maggots, etc.) and record them in the (Observations Outside Bucket) column.
- Photograph the bucket label to help organize photos later.
- Rotate the bucket 90 degrees to the left and take a photograph inside the ventilation tube. Focusing is hard in this photo but try to get an image that shows anything inside as best as possible. Make any observations (odor, substance, air pockets, organisms, liquid, color, etc.) and record in the (Observations Vent Tube) column.
- Before opening, spray the access port with 70% ethanol then wipe clean and allow to air dry. Repeat with 10% bleach.
- Place a paper towel on the floor in front of the access port.
- Sanitize the moisture probe, blue thermometer, and one spoon on the yellow lid once again, wipe clean with a paper towel, and place the tools on the paper towel in front of the bucket.
- Sanitize gloves.

Collecting/Observing/Photographing the Sample

- Be prepared to click "Save" after every data entry until the end of sampling for each bucket.
- Confirm ID numbers, enter a "Y" in the (Verified) column and record the sampler's initials in the spreadsheet (Sampler Initials) column.
- Open the access port and place the cap upside down on the paper towel.
- Make any observations (odor, substance, air pockets, organisms, liquid, color, etc.) and record in the (Visible Contents Access Port) column.
- Photograph the access port (Zoom 2x with flash)
- Insert thermometer probe and moisture probe (moisture % will read as a decimal) into the compost through the access port and record in the (Compost Temp C) and (Compost Moisture Percent) columns respectively.
- Remove the thermometer and moisture probe (taking care not to pull out excess compost; tip the bucket back if necessary) and place onto the paper towel.
- Place the small scale close to the sampling area, place the sample baggie on the scale and press TARE.
- Carefully open the baggie, then using a sample collection spoon or tongs, obtain 2.0 g of material from the access port and place into the open baggie. (Take care not to touch the inside of the baggie with the spoon.) Record the weight of the sample in (Sample Weight g) column.
- Remove an additional 0.1 and 0.3g of sample from the bucket and place into the pH sampling container. (See "pH testing" below).
- Take a clear photo of the sample (no flash and 1x zoom) being mindful to eliminate the glare from the sun on the baggie. Do this quickly to avoid condensation inside the baggie which can obscure the view of the sample and/or organisms.
- If organisms are present, use a ruler to measure the organism length and take another photo of the organism.
- Make all qualitative observations of the sample through the clear side of the baggie (opposite side of label with sample code). Record observations in the (Observed Sample Color) using the "Compost Color Chart" (Fig. S2), (Observed Sample Moisture), (Observed Sample Contents), (Observed Sample Smell), and (Observed Sample Organisms YN) columns of the spreadsheet, respectively. If there are organisms present, describe it in (Organism Type) and measure and record in Length Observed Organism mm).
- Shake the baggie to move the sample into a corner and then roll the baggie from the bottom to remove air and seal the top.

- Place the sample into the sample transportation bag to be delivered to the locked -20 degree Celsius freezer in the Applied Research and Development (ARD) building or the Genetics Core Lab Facility.
- Close the access port.

## pH Testing

- While one researcher is cleaning up the sampling area in preparation for another bucket, another researcher can conduct pH testing.
- Add enough water to saturate the compost in the pH sampling Tupperware (approximately 0.625 mL)
- Using the pH sample spoon, mix the water and sample for approximately 30 seconds and allow to settle for 1 minute.
- Set the spoon on a paper towel then dip a pH strip into the liquid portion.
- Compare the color of the strip to the color chart and record the pH in the (Compost pH) column.
- Throw the pH strip and compost/water mixture into the garbage and clean the Tupperware and spoon with ethanol and bleach.

## Workspace Sanitization after Sampling

- Wipe off excess compost from the thermometer, moisture probe, and the sample collection spoon with the paper towel that is sitting in front of the bucket and discard.
- Sweep up any debris and dispose of it into the garbage can.
- Sanitize the tools and sampling area on the tarp with ethanol and bleach.
- Place the thermometer and moisture probes onto the yellow lid and replace the sampling spoon into the black tray.
- Replace the bucket to its respective position in the crate according to the (Crate Location) column in the spreadsheet or the laminated image of bucket locations.
- Replace gloves with a new pair.
- Proceed to the next bucket.

## Clean Up

- After all buckets have been sampled, sanitize the entire workspace, including phones, computers, and tools.
- Do not rotate buckets in the crate on Tuesdays. Rotations only occur on Thursdays. Refer to the "Bucket Rotation Location Diagram" or the drawings/arrows on the floor of the crate.
- On the moisture sensor, remove the red alligator clip and turn off the multimeter.
- Be sure to unhighlight all rows from the day.
- Close ethanol and bleach spray bottle tips to avoid spills and evaporation.
- Place all clean/sanitized sampling instruments back into the black tray and store in the large black storage container.
- Place empty, placeholder buckets (used as seats) back into the workstation crate.
- Store the indoor/outdoor thermometer on the floor and the remaining two thermometers go on top of the bucket in crate location #20.
- Sweep the entire workstation area and dump debris into the garbage can.
- Close the workstation crate door and lock all latches.
- WASH HANDS!!
- Dispose of garbage.

## 1.5.2 Disposing of the Material

After 133 weeks, buckets were emptied into an actively managed, thermophilic composting pile at the Flagstaff EcoRanch, a local nonprofit sustainability education center in Flagstaff, Arizona. The material will further compost for one year before being spread on non-edible food crops. In preparation for future HEC projects conducted by our research team, all buckets underwent thorough cleaning, disinfection, maintenance, and storage.

## 1.6 Laboratory Procedures

### 1.6.1 DNA Extraction and Sequencing

Extractions were performed on HEC and FYWC samples using the Qiagen DNeasy® PowerSoil Pro Kit or the DNeasy® PowerLyzer® PowerSoil® Kit according to the manufacturer's protocol. Approximately 0.25 g was used for all composting material samples. Less was used for the bulking material because 0.25g was more than could fit into the bead tubes. For swab samples, approximately half the swab was removed and placed into the bead tube. The final elution volume was 100 μL. Multiple reagent blanks were taken through extraction, amplification, and sequencing to monitor laboratory and extraction reagent contamination.

Using the 515F (GTGYCAGCMGCCGCGGTAA)[4] and 806R (GGACTACNVGGGTWTCTAAT)[5] primers, the V4 region of the 16S rRNA gene was amplified from each of the samples and reagent blanks and prepared for sequencing on the Illumina MiSeq instrument similar to the protocol presented by[6]. Briefly, each sample was amplified in triplicate using 10 ul (0.8x) Platinum Hot Start PCR Master Mix (Invitrogen), 1 ul (0.2uM) 515f forward-barcoded primer, 1 ul (0.2 uM) 806r reverse primer (0.2 uM), 1.0 ul DNA and water to 25 ul.

Cycling conditions were as follows: 3 minutes at $94^{o}C$ followed by 35 cycles of $94^{o}C$ for 45 seconds, $50^{o}C$ for 1 minute and $72^{o}C$ for 1 minute 30 seconds and a final extension at $72^{o}C$ for 10 minutes. A negative template control was run for each of the barcoded primers to monitor for contamination. Each of the triplicate reactions were pooled and run on an agarose gel to confirm the appropriate size amplicon was produced. The separate libraries were quantified using a Qubit 4 Fluorometer and 240 ng of each library was combined into multiple tubes to be cleaned using the QIAquick PCR Purification Kit (Qiagen). After elution into 50 ul following the cleanup, all cleaned pools were combined into a single tube. The final pool was quantified using the KAPA Library Quantification Kit (Roche) and visualized using an Agilent 5200 Fragment Analyzer System for size quantification. A dilution of this final pool was then sequenced on an Illumina MiSeq instrument.

### 1.6.2 qPCR ASSAYS

#### 1.6.2.1 Primer Selection - uidA and *C. perfringens* group

*Escherichia coli* and *Enterococci* spp. are commonly used fecal indicator bacteria (FIB) used by regulatory agencies to monitor recreational waters[7,8]. Applying qPCR has shown to be more reliable for detecting FIB compared to culture-dependent methods[9]. Two PCR primer sets, uidA[10,11] and Cperf[12], were used to identify the presence of *E. coli,* a non-spore forming organism and a group of *Clostridium perfringens,* spore forming organisms respectively, as well as to compare their resiliency during mesophilic composting.

### 1.6.2.2 *in silico* PCR Screen

To test for sensitivity and specificity of published primers *in silico*, PCR primers were screened against 4385 reference bacterial genomes with the -search_pcr method in USEARCH v11.0.667 (ref. [13]), using the following options: "-strand both -maxdiffs 2 -minamp 70 -maxamp 1500". The probe was then queried with USEARCH against the amplicons with the following options: "-search_oligodb amplicons.fasta -db probe.fasta -strand both -userfields query+target+qstrand+diffs+tlo+thi+trowdots". The number and composition of predicted amplicons and probe were manually verified. The primers were deemed useful if they only amplified the species target of interest.

### 1.6.2.3 g-Block Design

To fully validate and optimize these PCR assays, a 300 Bp genomic block (Gene Fragment from Twist Biosciences) was ordered encoding these two species-specific targets (Fig. S3). Beyond assay characterization, we continued to use this synthetic GBlock as our positive control in this study. To enable us to generate more GBlock template in-house as needed, the GBlock was designed to include two unique universal insertion sequences (highlighted in green), each sequence flanking one end of the Gblock fragment. These strategically inserted universal sequences allow for easy propagation of the entire GBlock sequence by PCR amplification using universal primers (UT1 and UT2)[14].

### 1.6.2.4 Assay Optimization

All DNA extracts were screened for the presence of *E. coli* and *Clostridium* species DNA using two previously published TaqMan assays; *E.coli* and Cperf[10,12], which target the uidA gene and small subunit rRNA, respectively. Both assays were incorporated into a multiplexed PCR for the purpose of screening individual DNAs for both species in a single reaction.

The multiplexed TaqMan PCR was performed in triplicate 10µl reactions containing 1µl of undiluted compost DNA as template and the following reagents (given in final concentrations): 1× PCR buffer, 1x ROX dye, 5mM MgCl2, 0.2mM dNTPs, 0.04U/µl Platinum® Taq polymerase, 0.5µM of each primer (EcoliUID_F, EcoliUID_R, Cperf23S_F, and Cperf23S_R), and 0.125µM of each probe (EcoliUID_P labeled with FAM dye and Cperf23S_P labeled with VIC dye). Thermocycling for PCRs used the following conditions on an Applied Biosystems QuantStudio 7 flex instrument: 95°C for 8 min to release the polymerase antibody, followed by 45 cycles of 95°C for 5 sec and 60°C for 30 sec.

Each PCR run included a serial dilution that was performed using a synthetic GBlock control of known concentration (1ng/µl) and length (300bp) that contained both assay targets, which enabled the calculation of template copy number per 1µl in our stock solution ($3.041 \times 10^9$) and thus, estimates for each dilution. Serial dilutions spanned seven orders of magnitude ranging from $3.041 \times 10^7$ to 3.041 template copies and were performed in triplicate reactions.

### 1.6.3 Pan-enterovirus Assay

Post extraction, all HE samples were treated using Invitrogen's ezDNAse reagent prior to reverse transcription. The 10µL ezDNase reactions contained 8µL RNA input and were incubated for 2 minutes at 37ºC following manufacturer recommendations. The full volume of each DNase-treated sample was reverse transcribed using the SuperScriptTM IV First Strand Synthesis System (Invitrogen) following manufacturer recommendations. Initial primer annealing with random hexamers was performed with 10uL DNase-treated RNA input and incubated at 70ºC for 7 minutes. Following primer annealing, reverse transcription of RNA was performed by cycling at 23ºC for 10 minutes, 50ºC for 45 minutes, 55ºC for 15 minutes, and 80ºC for 10 minutes. All synthesized cDNA was treated with the included RNase H to remove residual RNA in the RNA/cDNA hybrids formed during the previous step. Following reverse transcription, samples were amplified using previously published pan-enterovirus primers targeting the 5' UTR region of the Enterovirus genome[15]. Reaction and cycling conditions for both the reverse transcription and pan-enterovirus PCR were the same as those found in[16].

### 1.6.4 BactQuant Assay

The BactQuant Assay (PMID: 22510143) was employed to detect a common bacterial gene, 16S rRNA, to determine relative abundance of bacteria in a sample[17]. The assay was performed as previously described[17] with modifications presented herein. 10uL reactions consisted of 1X PCR Buffer (ThermoFisher), 0.04U/uL Platinum Taq Polymerase (ThermoFisher), 5mM Magnesium chloride (ThermoFisher), 1X ROX reference dye FisherSci, 1.8uM of each primer (forward: CCTACGGGDGGCWGCA; reverse: GGACTACHVGGGTMTCTAATC), and 0.225uM of probe (6FAM-CCAGCAGCCGCGGTA-MGBNFQ). 1uL of ten-fold diluted genomic DNA in Tris-Tween (1M Tris-HCl, Tween-20) was submitted into reactions and qPCR conditions on the QuantStudio 7 Pro were: 5 min at 95°C, 15 sec at 95°C, 60 sec at 60°C for 40 cycles. 1043 genomic DNA samples were screened on this assay in triplicate alongside a 16S rRNA plasmid standard diluted $10e^8$ to $10e^2$ in Tris-Tween. Assay performance was evaluated following each run and standard qPCR metrics were achieved. 12 samples (0.01%) did not amplify or had a Ct value greater than 37.

Quantitative measurements of 16S rRNA in a sample were calculated using the standard curve performed on the respective run using the Design and Analysis software. Reversal of the ten-fold dilution was accomplished by multiplying the quantity by ten. Figure design and statistical tests were completed using Prism 10 software.

### 1.6.5 Culturing *E. coli* Using EC Medium with MUG

Half of the HEC samples (391 total from odd numbered TPs) were cultured using EC (*E. coli*) medium w/ 4-methylumbelliferyl-ß-D-glucuronide (MUG), a buffered lactose broth used for the fluorogenic detection of *E. coli*[18]. The uidA gene (qPCR assay used in the study) encodes for ß-glucuronidase and when present, cleaves MUG and fluoresces which can be detected with a UV light. The samples were previously frozen at -20°C for between one to two years. *E. coli* has shown to persist at subfreezing temperatures in Alaska at HE dumpsites[19].

To prepare the EC-MUG broth tubes, 37.1 g/L of powdered medium (Alpha Biosciences, Baltimore MD, USA) was dissolved in deionized water. Liquid medium (10 mL) was dispensed into 16x150 mm glass test tubes containing an inverted 6 x 50 mm Durham tube. The medium was sterilized by autoclaving at 121.1°C for 20 minutes and stored at 4°C until use. Quality assurance was performed via testing with *E. coli* K-12 strain as a positive control and *Klebsiella pneumoniae* as a negative control. The medium performed as expected with *E. coli* producing blue fluorescence and gas after growth at 44.5°C for 24 hours, while *K. pneumoniae* produced limited growth with no fluorescence and no gas production.

### 1.6.5.1 Preliminary Analysis

To determine the presence or absence of *E. coli* in the samples, 0.2 grams from each sample was placed into a broth tube, in duplicate, and incubated in a water bath at 44.5°C for 24 hours. The detection of gas bubbles in the Durham tubes and fluorogenic detection with UV light indicated the presence of *E. coli*. If one of or both tubes were positive, the sample ID was recorded, and a serial dilution was conducted at a later time to quantify the most probable number (MPN).

1.6.5.2 Serial Dilutions and MPN Calculation

Serial dilutions ($10^{-1}$, $10^{-2}$, $10^{-3}$, and $10^{-4}$) were conducted on positive samples. 0.5 grams of HEC was placed into 5mL of 1x PBS buffer and vortexed for 60 seconds to homogenize. One mL of each dilution was dispensed into 4 tubes containing EC-MUG media and incubated in a water bath at 44.5°C for 24 hours. The presence of gas bubbles and UV light detection were indicative of positive results for the presence of *E. coli*. The parameters for the calculations including the number of dilutions, number of tubes per dilution, and inoculum volume were entered into the US EPA's Most Probable Number (MPN) Calculator[20] to determine the MPN (Fig. S4).

# Supplementary Text 2: Data Analysis

## 2.1 Preliminary Analysis

Initially, samples were collected before and after rolling buckets (pre- and post-roll respectively). It was hypothesized that different microhabitats within each bucket (near the ventilation tube as opposed to the bottom surface) would contain different microbial communities[21]. Thirty-five weeks into the experiment, a preliminary study was conducted on Bucket 1 (B1) to test lab protocols, sampling methods, and determine if the microbiome was in fact changing over time. In July 2021, the V4 region of the 16S rRNA gene from 70 samples was sequenced on the Illumina MiSeq and using QIIME 2[22]. The paired differences between 70 weekly pre- and post-roll samples from B1 were compared (Fig. S5). Shannon diversity ($p = 0.634$), Faith's PD ($p = 0.634$), and observed features ($p = 0.907$) showed no significant difference between pre- and post-roll samples. Evenness ($p = 0.024$) was significant, but this may be due to aeration and homogenization or the location of or small volume of the sample removed. Similar findings were observed in pit latrines and dry CTs[23,24]. Consequently, pre-roll sample collection was terminated in favor of post-roll sample collection only, and previously collected pre-roll samples were omitted from further analyses, optimizing time and sequencing costs.

## 2.1 Bioinformatics Pipeline

Microbiome bioinformatics analysis was performed on the V4 region of 16S rRNA data using QIIME 2 v.2024.5[22]. Paired end sequences were imported using manifestor.py (https://github.com/gregcaporaso/fq-manifestor; commit: 0d36635) and quality control of the Amplicon Sequence Variants (ASVs) was performed using qiime dada2 denoise-paired[25]. The forward sequences were not trimmed (sequence length 251) and the reverse reads were trimmed at 250. This was separately applied to four sequencing runs in DADA2. The feature tables and representative sequences from these runs were merged respectively and used for the entire analysis.

A total of 1,397 samples were collected. The bac120_taxonomy_r214.tsv reference taxonomy and bac120_ssu_reps_r214.fna reference sequences from Genome Taxonomy Database (GTDB) release 214.1[26] were used to create a weighted stool classifier using qiime clawback[27,28]. The 16S V4 subregion was extracted from the reference sequences using the primers 515F_GTGYCAGCMGCCGCGGTAA[4] and 806R_GGACTACNVGGGTWTCTAAT[5]. An initial classifier was created using qiime feature-classifier fit-classifier-naive-bayes[27] with the reference taxonomy and extracted reads. Relevant sample IDs were fetched from Qiita [2] using redbiom[29] with the query strings "Stool" and "stool", these sample IDs were used alongside the context "Deblur_2021.09-Illumina-16S-V4-150nt-ac8c0b" to fetch sequences to be used as weights. The previously mentioned classifier was used to classify these weights using qiime feature-classifier classify-sklearn[27,30]. Using the reference taxonomy and reference sequences alongside the sample weights of interest we utilized qiime clawback generate-class-weights to create weights for the classifier[27,28]. The generated weights of interest were used to retrain our initial classifier using qiime feature-classifier fit-classifier-naive-bayes[27]. Lastly, taxonomic annotations were assigned using qiime feature-classifier classify-sklearn on the collected dataset[27,30] and taxa-barplots were created using qiime taxa barplot to track microbial diversity over 52 weeks.

All filtering steps were performed using qiime feature-table filter-samples. Initial filtering removed any ASVs with a frequency less than two to remove residual sequencing error. Additional filtering was performed to remove duplicate sample-ids and ASVs taxonomically annotated only to the domain level (since ASV's not taxonomically labeled to the phylum level are likely not 16S sequences and may be associated with human contamination, fungi, protists or archaea).

A phylogenetic tree was created with qiime phylogeny align-to-tree-mafft-fasttree[31,32] for use in phylogenetic diversity measures.

After performing rarefaction at an even sampling depth of 9,614 with q2-boots[33], we retained 92.13% of samples and 46.11% of reads, resulting in a final dataset of 1,018 samples and 9,787,052 reads. Application of rarefaction enabled us to consider all collected data, even though only 46.11% of reads were considered in each iteration. Sampling depth was confirmed to be relevant even at 9,614 using qiime diversity alpha-rarefaction curves based on Faith's Phylogenetic Diversity values plateauing[34,35]. One-hundred iterations of rarefaction were applied, and the resulting alpha and beta diversity metrics were averaged across iterations. Alpha diversity vectors were averaged using median and beta diversity distance matrices were averaged using medoid.

The feature table was collapsed to genus and order level using qiime taxa collapse. This collapses groups of features with the same taxonomic assignment and sums the feature abundances. Next, qiime composition ancombc compared the abundance of features in HE compared against the abundance of features in HEC at TP-52 (ref. [36]). These differentially abundant order and genus features were visualized using qiime composition da-barplot with a significance threshold of 0.05 q-value and a log fold change absolute value of 3.

The plugin q2-fmt was applied to determine if the microbiome of the HE had engrafted in the HEC[37]. All but one of the initial HE samples per bucket were grouped using qiime feature-table group serving as the "donor." The non-grouped HE sample served as the "recipient" and baseline to determine if the HEC microbiome had shifted compared to the "donors." qiime boots core-metrics was run with the same parameters as described above on the new grouped table. The resulting distance matrices were used as the diversity metric for qiime fmt engraftment. Raincloud plots (Extended Data Fig. 9) were used to track microbial engraftment from HE to TP-52 demonstrating the differences between HE and HEC at TP-52.

The final feature table was collapsed to the genus level and TPs 0, 55, and 60 were filtered out. qiime longitudinal feature-volatility[38] was applied to determine the change of microbial communities in HEC over one year and determine potentially relevant organisms.

All data analysis steps are outlined in full detail in the QIIME 2 Provenance Replay[39] reproducibility supplement in the Zenodo data archive (Data Availability).

## 2.2 Physicochemical Dynamics

### 2.2.1 Compost Temperature

Composting typically commences in the mesophilic (20-45°C) phase, progresses into a thermophilic (45-75°C) phase, and returns to the mesophilic phase. An optimal range of 50-65°C[40] is preferred to effectively destroy pathogens, accelerate decomposition, and kill weed seeds[41–44] and consequently, many composting regulations mandate maintaining temperatures of 55°C or higher for a specified duration to ensure pathogen elimination[45].

Small-scale composting systems like those examined in this experiment do not attain thermophilic temperatures unless they are heated externally[46,47]. Composting can proceed efficiently and reduce pathogens at lower temperatures provided longer durations are achieved[48–50]. This approach may conserve more carbon and nitrogen compared to thermophilic composting while achieving equivalent levels of compost maturity[40].

Compost temperatures in this study corresponded with seasonal and ambient temperatures (Extended Data Fig. 2a), similar to a previous study[21], decreasing to 6.9°C during winter months and increasing to 26.7°C in summer months. Variations in compost temperature between buckets on similar TPs may be attributed to their positions within the storage crate. Buckets located on the top row were exposed to warmer temperatures and direct sunlight compared to those in the middle or bottom rows.

### 2.2.2 Compost pH

The preferred composting pH range is between 6.5 and 8.0, although efficient composting can occur between 5.5 and 9.0 due to microbial diversity within the system[40,41,51]. The observed decrease in pH in this study (Extended Data Fig. 2b,c) is attributed to the hydrolysis of organic matter and the subsequent accumulation of organic acids and $CO_2$ dissolution[50,52]. The initial elevated pH in B5 and B7 may have resulted from ammonification under anaerobic conditions caused by high moisture content, despite weekly homogenization[53].

Typically, pH levels rise towards a more neutral pH by the end of the composting process[50]; however, all buckets, except B7, remained acidic. Insufficient nitrogen levels can hinder microbial activity, thereby reducing the rate of decomposition[40,54]. This results in the accumulation of organic acids, such as acetate, propionate, and butyrate, which produce foul odors. Foul odors were particularly strong in B5, B6, and B7, with B5 developing an earthy smell at TP-18, B6 at TP-15, and B7 at TP-18; however, a sour/ammonia odor returned in the buckets between TPs 28-40. *Clostridium* species have been associated with odor production in food composting plants[55] and could have contributed to the odors since their abundance increased over time as seen in the qPCR data (Extended Data Fig. 4). Partially decomposed organic matter may also undergo humification, leading to the formation of humic acids and further contribute to the acidity of the compost.

## 2.3 Ridgeline Plot Analysis

For each bucket, the distances between the first three (1-3) and last three (50-52) TPs of HEC were compared to each reference sample (FLWC, soil, BM, and HE) and plotted forming ridgeline plots. The change in position of the shaded area of TP 50-52, relative to the shaded area of TP 1-3(either towards or away from) illustrated increasing similarity or dissimilarity respectively (Fig. S6).

In all buckets except B1, TP 50-52 microbiomes increased in similarity between HEC and FLWC and all buckets became more similar to the soil samples over time. The microbiomes of B4, B5, B9, B11, B13, and B14 at TP 50-52 appeared increasingly dissimilar to the BM, likely due to progressive decomposition. In contrast, the remaining buckets exhibited no significant change, which could have resulted because BM constitutes a substantial proportion of the initial HEC, thereby predominantly contributing to the overall microbiome composition. The greatest shift in all buckets is evidenced by the shift of TPs 50-52 away from the HE samples, indicating that the microbiomes of the final HEC samples are most dissimilar to the HE microbiome and therefore no longer representative of HE.

## 2.4 Feature Volatility

Feature volatility (qiime longitudinal feature-volatility) can predict temporal relationships based on feature abundance[38] and can be used to explore potential biomarkers[56]. The top 10 most important features (rank # follows the taxon in parentheses) revealed significant ($p<0.0001$, $r^2 = 0.937$) patterns and ecological interactions of the microbial succession and functional dynamics indicating a transition in the composting systems between TP-20 and TP-35. Taxa that decrease over time are low to undetectable by TP-35, those that increase over time begin doing so around TP-20 (Extended Data Fig. 8).

*Hyphomicrobium* (#1) and *Acidimicrobiales* (#3) displayed a progressive increase throughout the sampling period. Increasing abundances of these taxa may co-occur due to their complementary roles in nutrient cycling[57–59] and organic pollutant degradation, which subsequently reduces odors[60,61]. The observed increases in these taxa may be attributed to consistent aeration of the buckets, which raises dissolved oxygen content (DOC)[62].

In contrast, *Arthrobacter_D* (#2) and *Pseudomonas_E* (#4) show a quick and steep decline until TP-5, becoming nearly undetectable by TP-25 and *Sphingobacterium* (#6) displays a steep decrease until TP-15. This suggests their involvement in the early stages of composting, playing crucial roles in the degradation of organic matter[63–65] and nutrient cycling[66]. *Arthrobacter* produces siderophores (iron-chelating compounds) possibly alleviating the concerns of heavy metals in HEC[63], is capable of co-metabolizing polychlorinated biphenyls (PCBs)[67], and demonstrates antifungal activity[68,69]. *Pseudomonas* is active in the mesophilic phase utilizing easily degradable organic material[70] and has been shown to degrade herbicides, pharmaceuticals[71], lignocellulose[72], and chitin[54]. Together with *Sphingobacterium*, it can degrade soils and wastewater contaminated with oil[73,74].

The family Pirellulaceae (#10) decreased more gradually until TP-35 with large fluctuations in some buckets. This taxon is usually associated with later stages of composting, contributing to the degradation of cellulose and hemicellulose and accelerating compost maturation[65,75].

B7 and B11 show sporadic, significant increases in the genera *JAAYYF01* (#7) and *Vulgatibacter* (#9), as well as the family Trueperaceae (#8), which contains *JAAYYF01* (NCBI reference - MAG: Trueperaceae bacterium isolate BROCD029; Accession# PRJNA647942 (ref. [76]). B6 (spouse of B7) exhibited an increase in *JAAYYF01* approximately three weeks after the increase observed in B7. B11 demonstrated a large spike in *Vulgatibacter* between TPs 19-35 and has been shown to degrade lignocellulose[77]. *CADCWL01* (NCBI reference - uncultured Thermomicrobiales bacterium isolate AVDCRST_MAG19), found in soil crusts[78], was most prevalent in B1 (decreasing before TP-30) and B4 (with sporadic fluctuations throughout). Trueperaceae is associated with thermophilic composting[56] and its member *JAAYYF01* contributes to anaerobic digestion in WWTPs and activated sludge samples as well as forming biofilms, possibly contributing to bacterial dynamics or stabilization during the compost process[76,79].

Bucket-specific patterns may be attributed to specific ecological conditions or interactions with other microbial taxa underscoring the importance of considering both spatial and temporal dynamics within diverse composting systems to optimize microbial succession and functional efficiency. Identifying taxa with significant bioremediation capabilities can help mitigate concerns about HEC composting and offers potential for influencing regulations regarding the persistence of pharmaceuticals (e.g., antibiotics, endocrine-disrupting chemicals), pathogens, heavy metals, and persistent pollutants (e.g., PFAS, PCBs) associated with HEC.

## 2.5 q2-FMT

Fecal Microbiota Transplants (FMTs) are a US Food and Drug Administration (FDA) approved treatment for recurrent *Clostridium difficile* infections[80], where a "donor" microbiome is transferred into a "recipient." FMTs are also being researched for various human health applications including enhancing cancer treatment outcomes[81] and digestive issues related to autism[82,83]. However, there is no standardized method for assessing the extent of microbiome engraftment.

New methods have recently been defined for assessing engraftment extent following FMT[37]. When these methods are applied to a compost system, the HE is viewed as the donor and BM is viewed as the recipient. Unlike an FMT, in a composting system we expect the HEC to become less like the fecal donor over time. If the microbiome of HE engrafts in the composting system, the microbiome of HEC should resemble that of the donor. The q2-FMT plugin[37] allows the user to track longitudinal changes for each bucket highlighting and tracking their individual movement within the system.

The patterns observed in the ridgeline plots complement the findings from the q2-FMT analysis (Extended Data Fig. 9). Specifically, the ridgeline plots indicate that HEC samples at TP 50-52 show increased similarity to reference samples while becoming more dissimilar to HE over time. This mirrors the q2-FMT results, which demonstrate a progressive divergence of HEC samples from the original HE microbiome, with Unweighted UniFrac distances increasing significantly from TP-0 to later TPs. Both analyses underscore a clear transition of the microbiome in the composting system, highlighting the shift from the initial HE profile towards a composition more akin to soil and FLWC, and away from its original "donor" microbiome.

## 2.6 ANCOMBC

ANCOMBC is a statistical method employed to identify differentially abundant taxa across sample groups. A comparison was made between TP-52 and the original HE samples to determine which taxa were enriched (increased) and depleted (decreased) over time (Fig. S7). At TP-1, the microbiome was dominated by genera typically associated with the human gut, including *Bacteroides*, *Blautia*, and *Faecalibacterium*. Over the course of 1 year, these taxa decreased significantly, while taxa associated with soil or other environmental habitats became more prominent. For instance, *Rhodococcus*, a genus widely found in various environments and known for its role in bioremediation[84] and WWTPs[85], exhibited a log fold change of +4.59 compared to HE samples. Conversely, *Faecalibacterium* showed a log fold change of -8.86. Similar decreases in human gut-associated bacteria have been observed in other HEC experiments[86–88].

*Rhodococcus*, *Devosia*, Rhizobiaceae, *Rhodanobacter*, and Chitinophagaceae are associated with key composting functions and play crucial roles in bioremediation, pollutant degradation, plant growth promotion, compost maturation, nitrogen fixation and nutrient cycling[72,84,89–94]. Chitinophagaceae, of which *Arachidcoccus* is a member, contributes to the decomposition of chitin and complex polysaccharides, further aiding in the breakdown of organic matter[95]. These taxa can facilitate the transition from a HE microbiome to a compost microbiome by enhancing nutrient cycling and organic matter degradation.

Collectively, the analysis with ridgeline plots, q2-FMT, and ANCOMBC demonstrates that taxa associated with the human gut have been significantly reduced or eliminated through composting, resulting in a microbiome that increasingly resembles that of soil and finished compost.

## 2.7 Microbiome Transition Patterns by Spouse

Cohabiting spouses (B2/B3 and B6/B7) were compared (Extended Data Fig. 7a). B2 and B3 begin and end in similar locations along Axis 1, following a trajectory that aligns with the average. One HE sample from B3 was an outlier, likely due to contamination, mislabeling, or sequencing errors. However, when comparing distances between this sample and other HE samples we were not able to statistically confirm this sample was an outlier using 1.5xIQR.

One interesting observation in taxa bar plots of B2 and B3 occurred between TP-21 and TP-22 (Extended Data Fig. 10) *Rhodanobacter* increased from 0.62% to 14.44% and *Arachidoccus* increased from 0.65% to 23.4% in B2. In B3, *Rhodanobacter* increased from 2.93% (TP-21) to 33.0% (TP-22) and *Arachidicoccus* increased from 1.18% (TP-22) to 16.74% (TP-23). In B12 (unrelated to B2 and B3) *Rhodanobacter* increased from 0.35% (TP-30) to 22.5% (TP-32) and *Arachidicoccus* increased from 0.77% (TP-31) to 9.24% (TP-32). No other buckets exhibited these patterns; however, minor fluctuations were observed, although less pronounced.

*Rhodanobacter* possesses antifungal activity[93], degrades pollutants, and removes NH3-N during composting[96]. It can also decompose lignocellulose[72] which can accelerate the maturation process of the compost[97]. *Arachidococcus*, improved plant growth in rhizospheric soils[94] however, displayed plant inhibition under the presence of high concentrations of phenylalanine when feather-based compost was applied to soils[98].

Furthermore, concurrent patterns of the genus *Limiplasma*, associated with the chicken gut microbiome[99] and *Clostridium* were observed. These participants run a small farm with chickens and could have transferred DNA through BM. B3 displayed an increase of *Limiplasma* from 1.00% (TP-44) to 16.44% (TP-45) and *Clostridium* from 0.18% (TP-44) to 8.21% (TP-45). These genera exhibit concurrent patterns over the sampling period with a prominent and quick spike in relative abundance; *Limiplasma* 0.722% (TP-36) to 13.15% (TP-37) and *Clostridium* 0.64% (TP-36) to 9.73% (TP-37). All other buckets tended to follow similar patterns across the sampling period except B3 and B13.

The literature on *Limiplasma* in composting is limited, however, a previous study[72] found that it may increase volatile fatty acids (VFAs) production, thereby creating readily available carbon sources for denitrifying bacteria. Relative abundances of *Clostridium* may have increased due to a loss of free air space (due to compaction of HEC), persistent mesophilic temperatures, and readily available nutrients through decomposition or as metabolic byproducts from other taxa[100].

B6 and B7 began their transition in similar locations close to HE; however, B7 quickly stalled in its transition, while B6 rapidly advanced its successional trajectory after TP-31 (Extended Data Fig. 7b). B5, B6, and B7 required intervention with the addition of bulking material (pine shavings) to reduce moisture content and increase the C:N ratio, thus reducing foul odors. This may be attributed to several factors, including improper use of the composting toilets by participants (e.g., lack of urine diversion or insufficient addition of bulking material), dietary influences, toilet usage patterns, deviations in composting variables from optimal ranges, or shifts in microbial community dynamics during the composting process. Further investigation is required to elucidate the precise cause. HE samples from B7 exhibited high relative abundances of *Blautia* (HE1=71.12%; HE2=21.96%; HE3=25.02%) and *Faecalibacterium* (HE1=2.34%; HE2=23.25%; HE3=21.95%) compared to B6, which had abundances of *Blautia* (HE1=16.43%; HE2=8.55%; HE3=9.98%) and *Faecalibacterium* (HE1=9.92%; HE2=12.16%; HE3=11.49%).

Both genera, commonly associated with the human gut[101] and expected to be identified in this study, have been observed in other composting studies[86,87]. *Blautia* is known for its anti-inflammatory responses[102], reduction of visceral fat[103], and probiotic properties[104]. *Faecalibacterium*, an abundant commensal bacterium, produces short-chain fatty acids through the fermentation of dietary fibers[105], and is linked to anti-inflammatory effects[102] and immunity[106].

## 2.8 qPCR Analyses

### 2.8.1 *E. coli* and *C. perf* Assays

A multiplexed TaqMan assay consistently detected *E. coli* and *Clostridium* species down to $3.041 \times 10^1$ gene copies, the lower limit of detection (LOD), for both targets. The linear range of the assay (i.e., the quantifiable range) was calculated at $3.041 \times 10^{\wedge}2$ template copies with $R^2$ values of 0.999 for both targets and assay efficiencies of 95.29 and 94.69 for *E.coli* and Cperf, respectively. Normalized (Extended Data Fig. 4) and non-normalized (Fig. S8) results are presented.

qPCR techniques have been applied to detect fecal indicator bacteria in environmental samples[9,107,108]. However, these methods can be inhibited by soil components that interfere with DNA polymerase, primer extension, or fluorescence[109,110]. Additionally, dead cells or residual DNA can affect qPCR analyses; however, during composting, the rapid turnover of organic material and degradation of dead cells minimizes this issue[111]. Reliable quantification of *E. coli* O157 has been achieved in environmental samples[112], and successful investigations have been conducted on *E. coli* in dairy manure under various environmental conditions[11].

The use of fecal indicators is crucial for managing the safety of HEC systems[12,23,113,114] and, when combined with amplicon sequencing, can enhance pathogen profiling[23]. Concerns with CTs include their failure to meet International Plumbing Codes[115,116] and composting definitions and physicochemical requirements. Despite this, CTs are marketed as producing safe and sanitized compost[117]. To address these concerns, qPCR was employed to identify the presence of *E. coli* (non-spore former) and *C. perfringens* (spore former) in each bucket at each TP.

### 2.8.2 Pan-enterovirus Assay

A pan-enterovirus assay[16] was only performed on pooled HE samples. No bands were present in gels at 500bp during the first screen indicating there were no enteroviruses detected.

### 2.8.3 BactQuant Assay

BactQuant is a broad-coverage quantitative real-time PCR assay designed to quantify 16S rRNA gene copy numbers and estimate bacterial loads within complex microbial communities (Extended Data Fig. 6)[17]. There was no significant correlation between copy numbers and time ($\rho = -0.03$, $p = 0.35$). Bacterial load decreased over time in B1, B3, B5, B6, B7, B10, B12, and B14, while it increased in B4, B9, B11, and B15; in contrast, bacterial load remained stable in B2, B13, and B16. Microbial biomass generally decreases over time during composting due to diminishing nutrient availability[118,119], however, bucket-specific conditions may have influenced the opposite patterns. For instance, there was an observed increase in biomass in soils fertilized with cattle manure, suggesting that nutrient availability might contribute to increased bacterial load[120].

## 2.9 Culturing *E. coli*

The regrowth capability of *E. coli* was determined using EC broth with MUG[18]. The presence was determined by the detection of gas bubbles in Durham tubes and fluorogenic detection with UV light (Extended Data Fig. 5). If one of or both tubes were positive, the sample ID was recorded and a serial dilution was conducted to quantify the most probable number (MPN) using the US EPA's MPN calculator[20].

The initial experimental design omitted culturing *E. coli* due to logistical constraints; however, we retroactively cultured the samples to assess growth which may have influenced the culturing results. Nonetheless, all buckets exhibited a decline in MPN over time ($\rho = -0.546$, $p < .00001$), with MPN levels becoming undetectable by TP-25 in all buckets, supporting the validity of our culturing protocol, which aligned with a previous study[121]. *E. coli*, being a facultative anaerobe, can grow in both aerobic and anaerobic conditions, and its lower numbers may be attributed to its sensitivity to microbial competition[122].

Detection and quantification of *E. coli* using qPCR after TP-25 likely reflects the method's sensitivity in detecting viable but nonculturable (VBNC) or non-viable *E. coli* DNA[11]. B16 exhibited no growth during culturing despite high qPCR copy numbers between TPs 3-7. Interestingly, B6 displayed the highest qPCR copy numbers of *E. coli* but recorded the lowest MPN values (excluding B16), with no growth of *E. coli* observed after TP-7.

The reduction of *E. coli* after 25 weeks may have been influenced by taxa identified in the feature volatility plots as changes in these taxa also appeared to fluctuate around 25 weeks. However, further investigations into their interactions would be required.

# Supplementary Figures

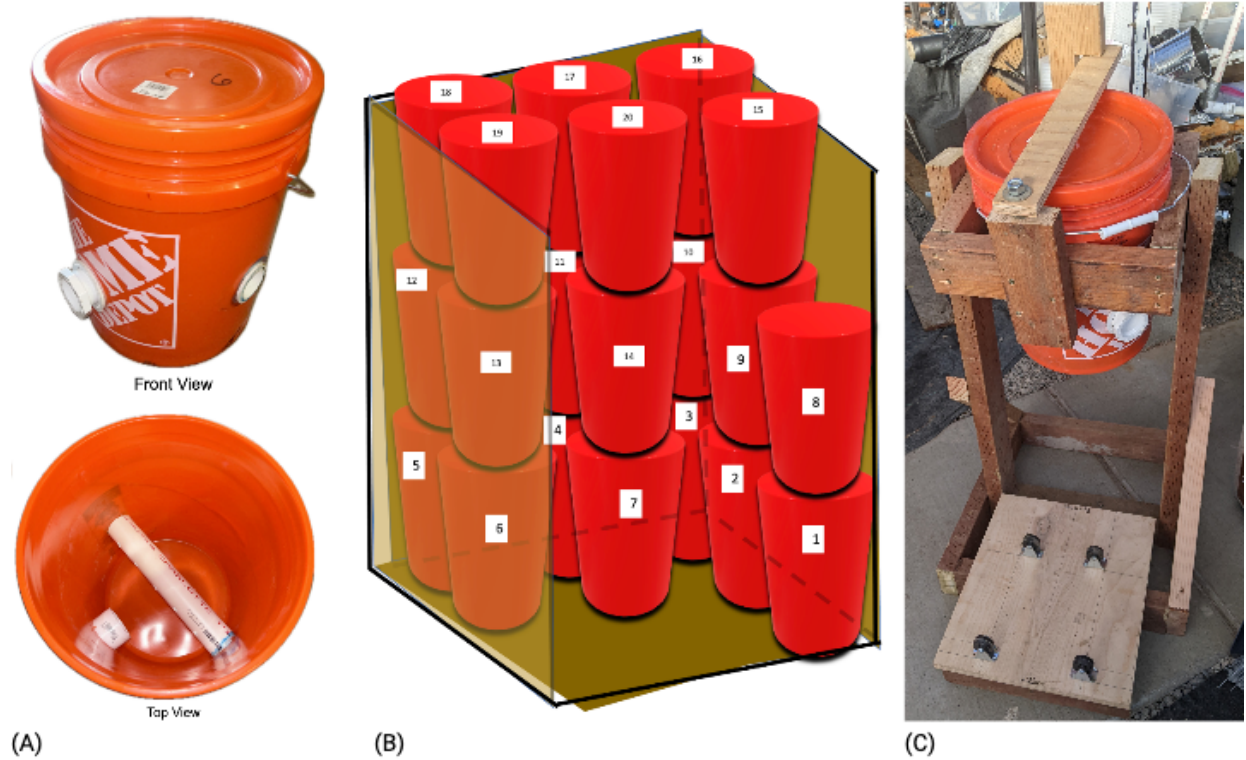

**Fig. S1** (A) Compost bucket design. (B) Rotation guide. Each week, buckets were advanced to the next numerical position (e.g., 1 → 2; 2 → 3...), with bucket 20 returning to position 1. (C) Bucket roller used to homogenize and aerate the HEC. Created in BioRender.

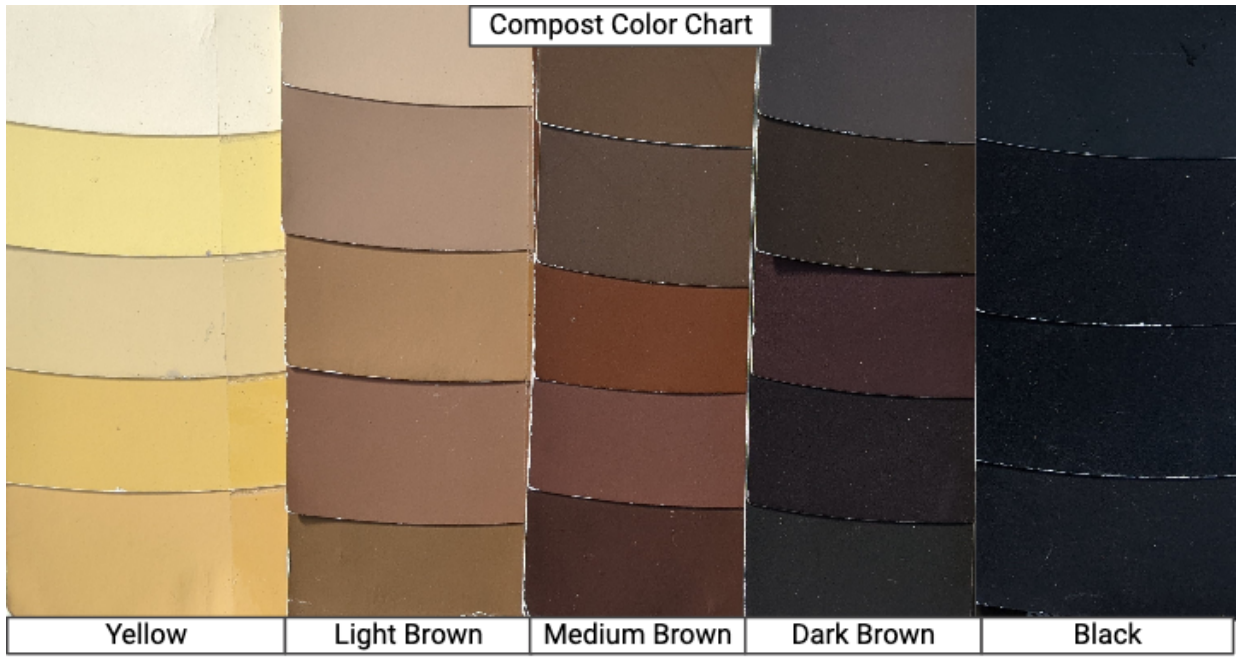

**Fig. S2** Compost color chart using paint swatches. Created in BioRender.

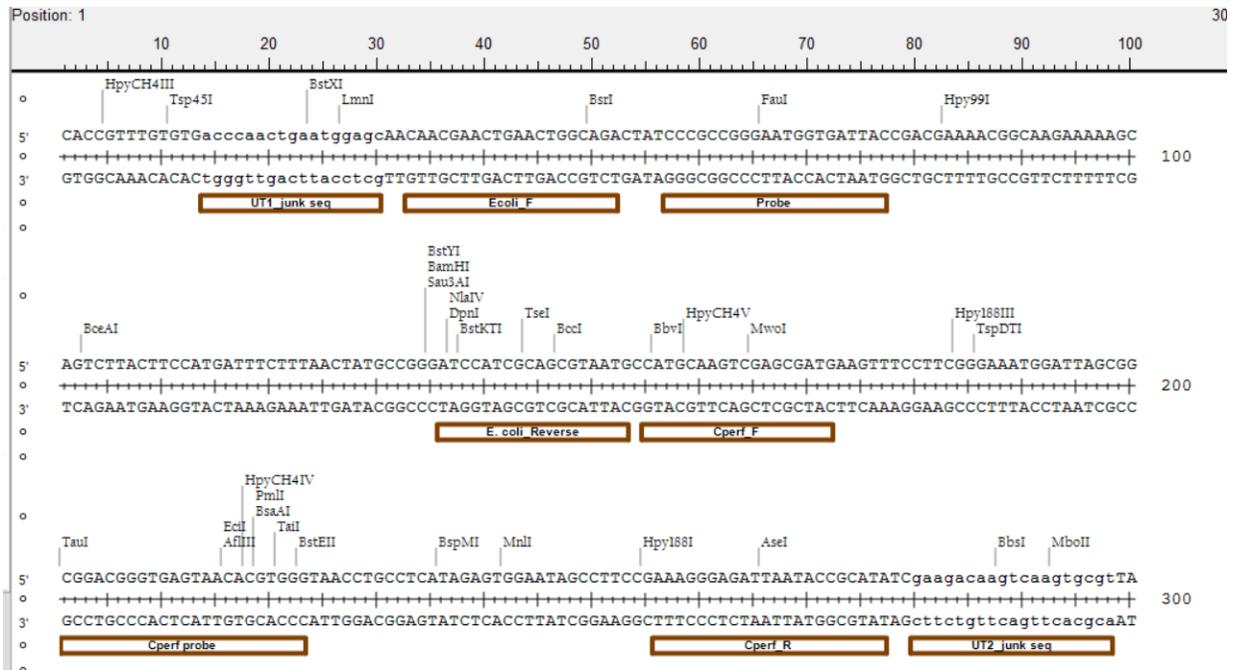

**Fig. S3** E.coli.Cperf.GBlock300bp. Universal primers (UT1 and UT2) that enable the propagation of the entire GBlock sequence by PCR amplification[14]. UT1 primer: acccaactgaatgggagc; UT2 primer: acgcacttgacttgtcttc.

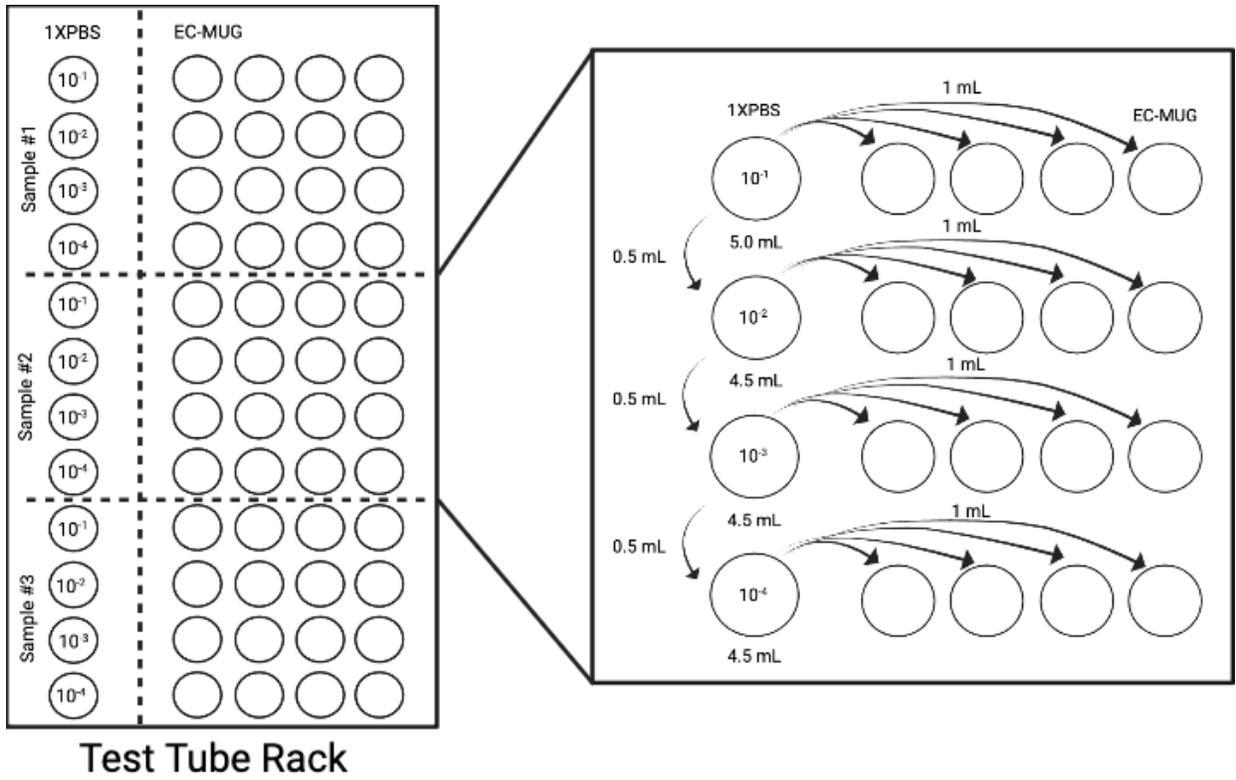

**Fig. S4** Serial dilution preparation and pattern of decreasing concentration solutions from an original sample.

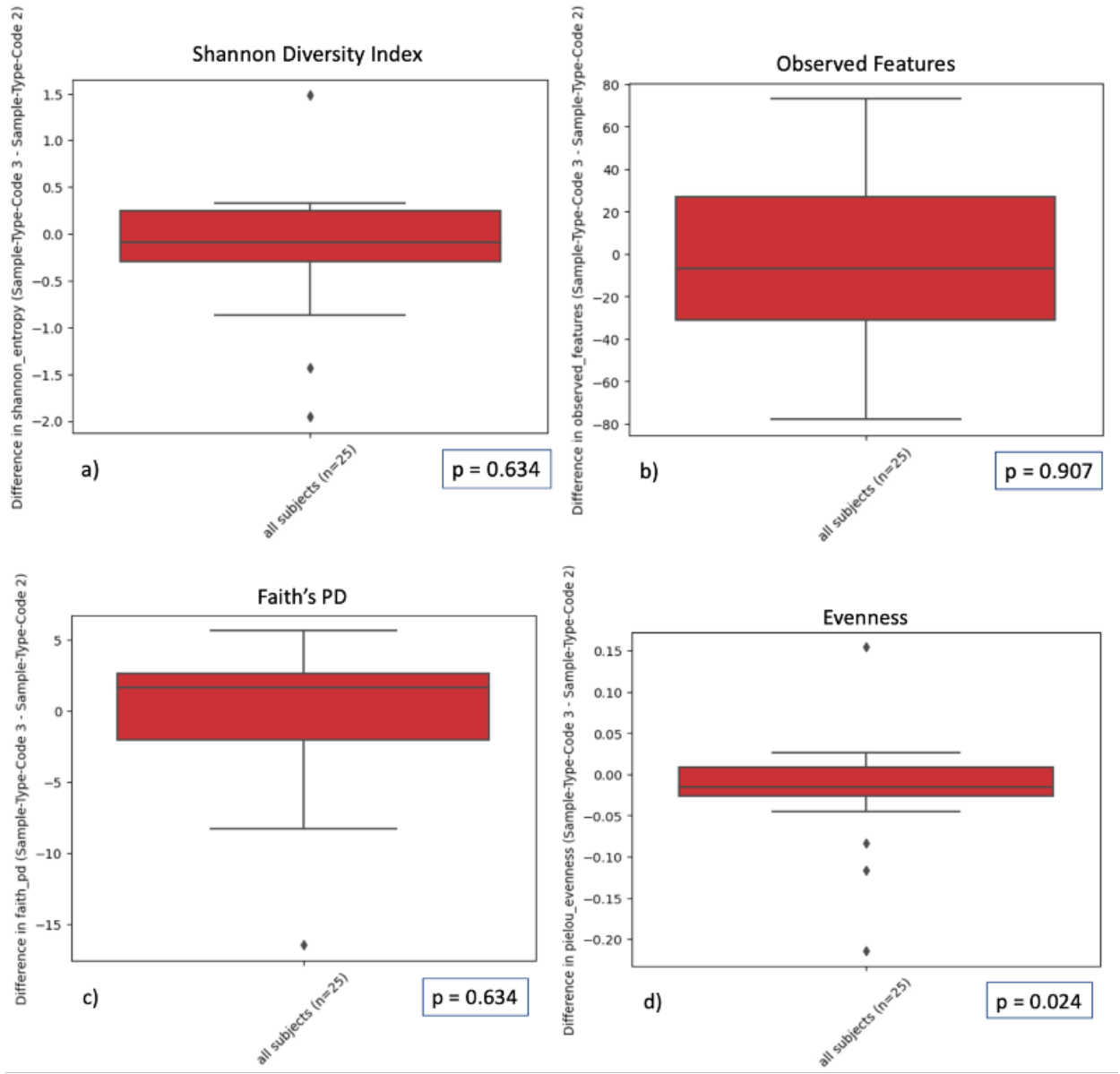

**Fig. S5** Pairwise differences to measure alpha diversity between weekly pre- and post-roll samples. Evenness (d) was significant but may be due to material that was not homogenized well, the location the sample was removed from inside the bucket, small sample size, or a false positive (after Bonferroni adjustment for multiple-comparisons, this p-value would be > 0.05

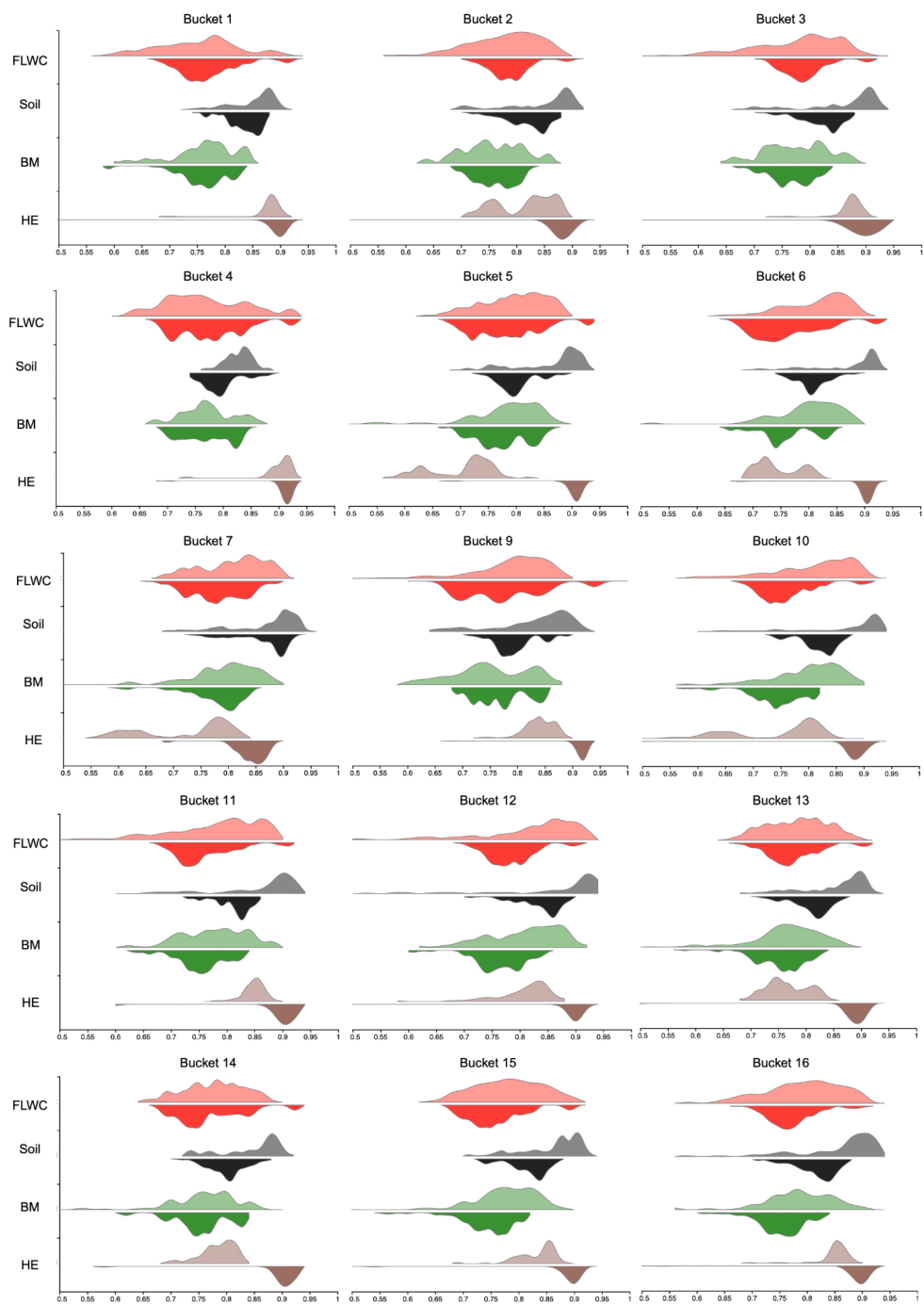

**Fig. S6**. The distances between the first three (1-3) and last three (50-52) TPs of HEC compared to each reference sample listed on the y-axis were plotted. Red represents distances between HEC and FLWC; black, between HEC and soil; green, between HEC and BM; brown, between HEC and HE. Lighter shades (top half) represent TPs 1-3 and darker shades represent TPs 50-52. For example, the distances between HEC TP-1 samples from B1 were compared to every FLWC sample (red) and plotted, followed by TP-2 and TP-3. All data points are found within the light red shaded area. This was repeated for each bucket and reference sample, and then repeated for TPs 50-52 in the bottom plots. A shift to the left along the x-axis of the TP 50-52 clouds indicates that the HEC becomes more similar to that reference sample. Conversely, a shift to the right indicates that the HEC becomes more dissimilar.

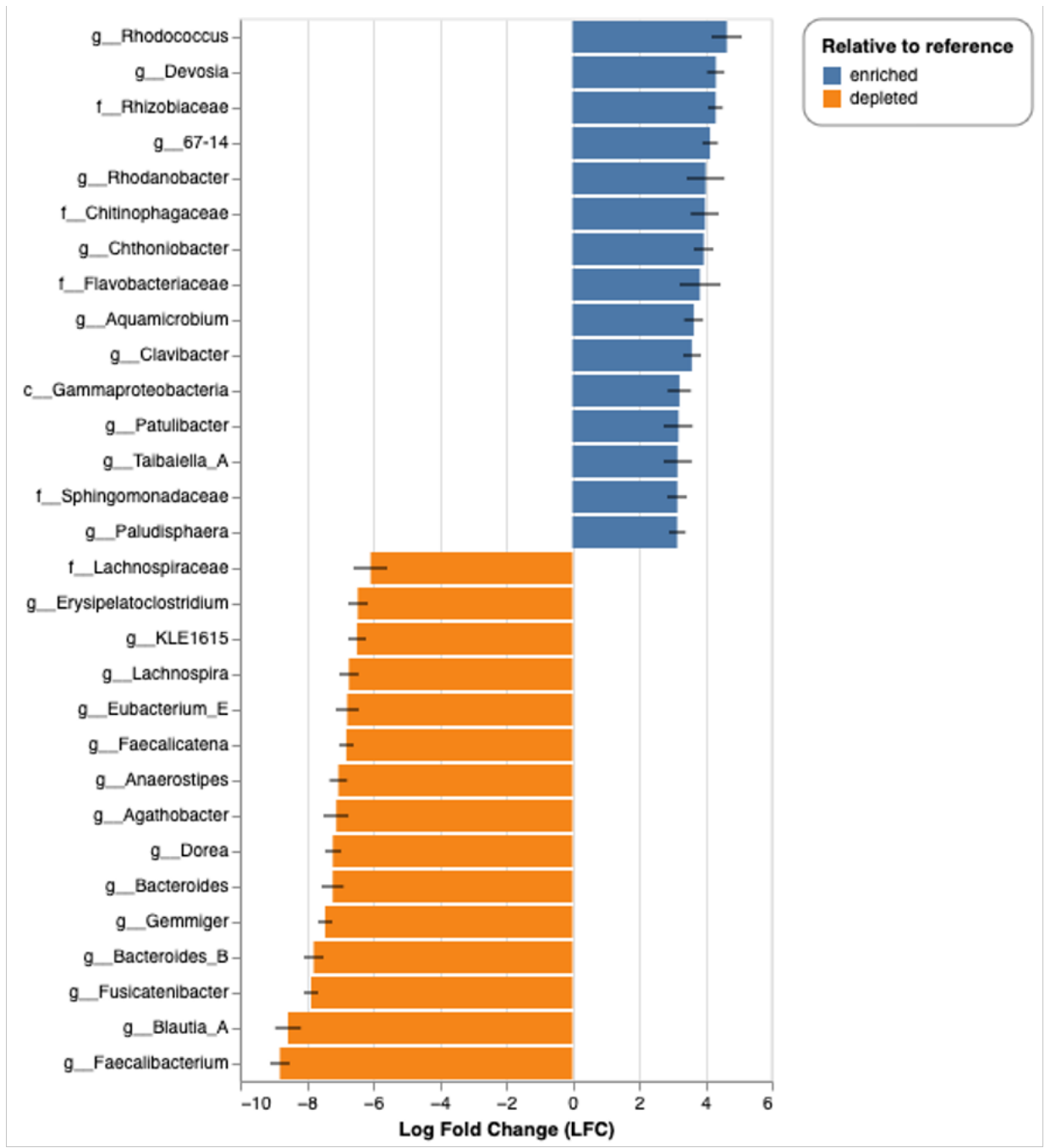

**Fig. S7** The Analysis of the Composition of Microbiomes with Bias Control (ANCOMBC) comparing TP-52 of HEC to HE showing the top 15 enriched and depleted genera that exhibited a log fold change greater than three. Created in BioRender.

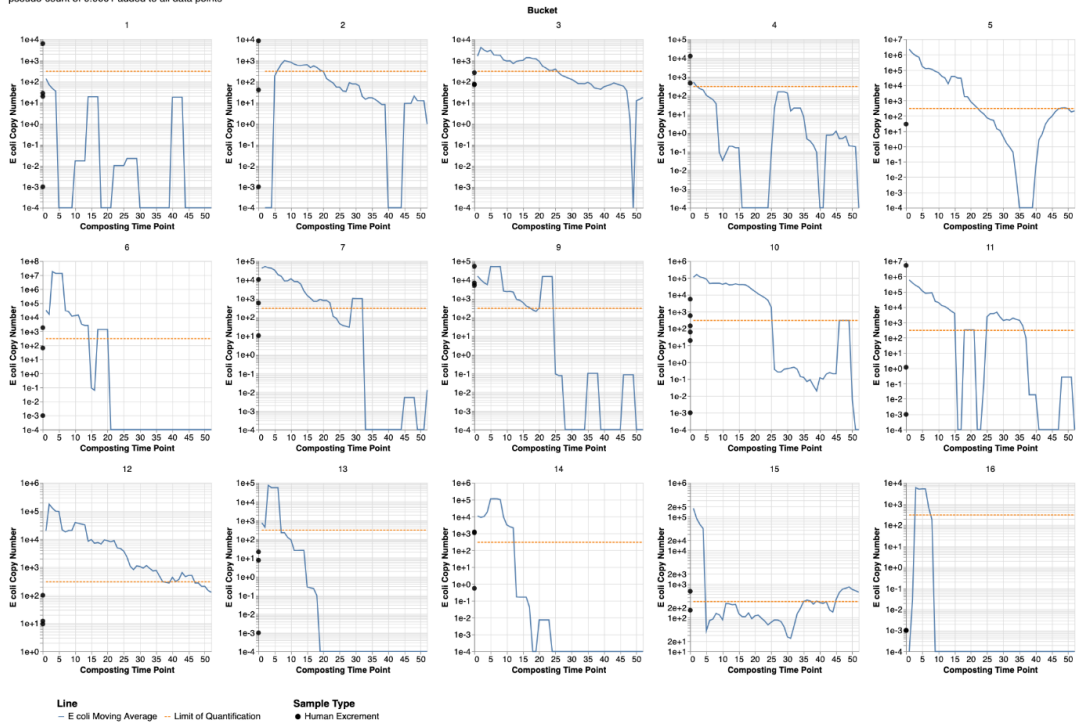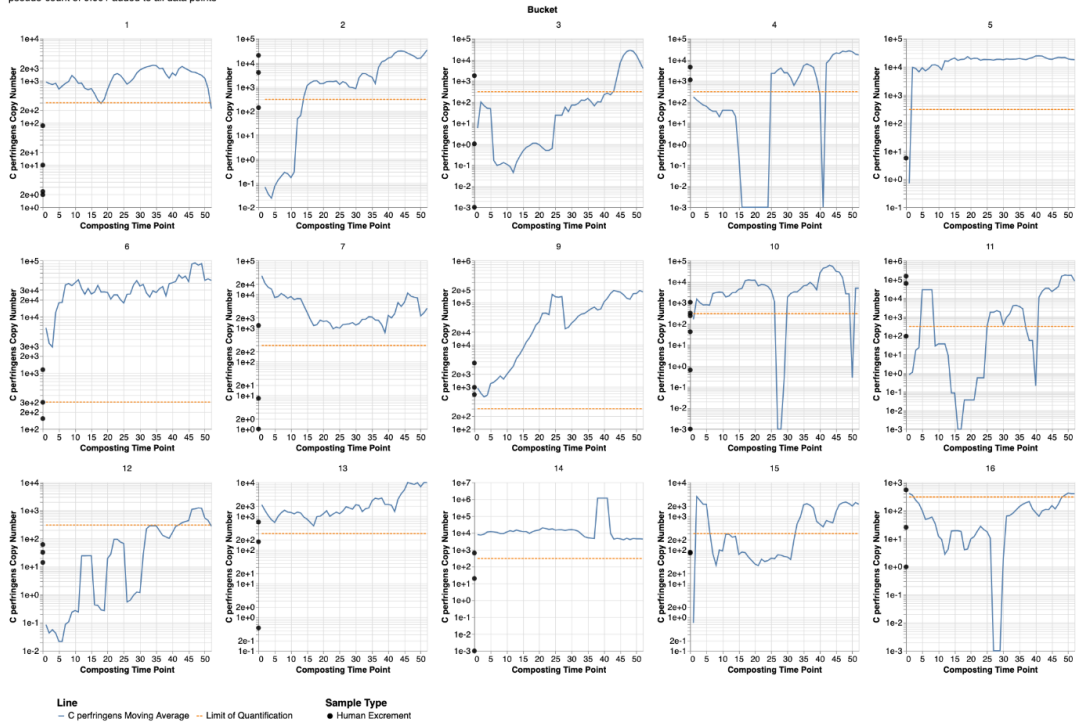

**Fig. S8** Non-normalized qPCR results for all buckets. (a) *E. coli* and (b) *C. perfringens*. Created in BioRender.

# Supplementary Tables

**Table S1** Participant Health Data survey.

| |
|---|
| Project Title: Bridging the Gap between Gut and Soil Microbiomes - Microbial Succession of Human Excrement in Composting Toilets |
| Principal Investigator: Greg Caporaso |
| Primary Researcher: Jeff Meilander |
| IRB: 1773199-1 |
| Please answer the following questions. You may provide as much or as little information as you feel comfortable with. |

Name _______________________________ DOB ______ Age ______ Height __________ Weight __________

Do you have any diseases? Yes ______ No ______

What disease(s) do you have? ________________________________________________________________

What is the frequency of your physical activity?

0-1 time/week ______ 2-3 times/week ______ 4-5 times/week ______ 6-7 times/week ______

What is the average length of time you spend exercising for each occurrence in the week?

0-15 min ______ 16-30 min ______ 31-45 min ______ 46-60 min ______ 61+ min ______

How would you characterize your diet?

Meat heavy (meat at most or every meal) ______

Meat light (some meat; not at every meal or not every day) ______

Vegetarian ______

Vegan ______

Pescatarian ______

Raw Food Diet ______

Keto Diet ______

Paleo Diet ______

Atkins Diet ______

Other (please describe) ________________________________________________________________

How would you rate your weekly stress levels?

None ______ Low ______ Medium ______ High ______

I have read and signed the consent form and I am willingly providing the information above for the research study.

_______________________________________________

Signature                Date

**Table S2** Participant and Composting Toilet Information.

| Participant | Bucket | Toilet System | Bulking Material | Notes |
|---|---|---|---|---|
| 1 | 1 | SunMar Excel NE (non electric) - self contained batch | Straw and sawdust | Replicates 1,12,16 |
| 2 | 2 | Homemade - 19L bucket | Pine shavings | Spouse of #3 |
| 3 | 3 | Homemade - 19L bucket | Pine shavings | Spouse of #2 |
| 4 | 4 | 3 Barrel System - continuous | Sycamore shavings | |
| 5 | 5 | Nature's Head - urine diverting self contained batch | Coconut coir | |
| 6 | 6 | Homemade - 19L bucket | Sawdust | Spouse of #7 |
| 7 | 7 | Homemade - 19L bucket | Sawdust | Spouse of #6 |
| 8 | 8 | Homemade - 19L bucket | Native grass and straw | Discarded at TP-2 due to no urine diversion by participant |
| 9 | 9 | Homemade - 19L bucket | Coconut coir | |
| 10/11 | 10 | Homemade - 19L bucket | Mesquite shavings | Shared by 2 participants |
| 12 | 11 | Homemade - 19L bucket | Sawdust | |
| 1 | 12 | Homemade - 19L bucket | Sawdust | Replicates 1,12,16 |
| 13 | 13 | Nature's Head - urine diverting self contained batch | Coconut coir | Replicates 13,14,15 |
| 13 | 14 | Nature's Head - urine diverting self contained batch | Coconut coir | Replicates 13,14,15 |
| 13 | 15 | Nature's Head - urine diverting self contained batch | Coconut coir | Replicates 13,14,15 |
| 1 | 16 | Homemade - 19L bucket | Pine shavings | Replicates 1,12,16 |